\documentstyle[sprocl,twoside,graphicx,amssymb,mycite]{article}

\newcommand{\bbox}[1]{\mbox{\boldmath$#1$}}
\newcommand{\diag}{\mathop{\rm diag}}
\newcommand{\text}[1]{{\rm #1}}
\newcommand{\case}[2]{{\textstyle\frac{#1}{#2}}}
\newcommand{\real}{\mathop{\rm Re}}
\newcommand{\Det}{\mathop{\rm Det}}
\newcommand{\Tr}{\mathop{\rm Tr}}
\newcommand{\tr}{\mathop{\rm tr}}
\newcommand{\transfer}{{\Bbb T}}
\newcommand{\bra}[1]{{\langle #1 |}}
\newcommand{\ket}[1]{{| #1 \rangle }}
\newcommand{\braket}[2]{{\langle #1 | #2 \rangle }}
\newcommand{\bracket}[3]{{\langle #1 | #2 | #3 \rangle }}

\begin{document}

\sloppy

\textheight=6.98truein   
\thispagestyle{empty}
\vspace*{0.85in}
\centerline{\LARGE\bf CHAPTER 39}  

\vspace*{0.21in}
\noindent \rule[0.5pt]{4.5in}{2pt} \\
\noindent \rule[8.5pt]{4.5in}{0.5pt}
\vspace*{0.25in}

\begin{center}
\huge\bf
Uses of Effective Field Theory in Lattice~QCD
\end{center}

\vspace{2cm}

\begin{center}
{\Large\bf Andreas S. Kronfeld}
\vspace{0.7cm}
\end{center}

\newpage

\pagestyle{myheadings}  
\markboth{\small\em Handbook of QCD / Volume 4}{\small\em Lattice QCD}

\title{\uppercase{Uses of Effective Field Theory in Lattice QCD}}

\author{Andreas S. KRONFELD}

\address{Theoretical Physics Department,
Fermi National Accelerator Laboratory \thanks{%
Fermilab is operated by Universities Research Association Inc.,
under contract with the U.S.\ Department of Energy.} \\
Batavia, Illinois, USA}

\maketitle

\abstracts{
Several physical problems in particle physics, nuclear physics, and
astrophysics require information from non-perturbative QCD to gain a
full understanding.
In some cases the most reliable technique for quantitative results is to
carry out large-scale numerical calculations in lattice gauge theory.
As in any numerical technique, there are several sources of uncertainty.
This chapter explains how effective field theories are used to keep
them under control and, then, obtain a sensible error bar.
After a short survey of the numerical technique, we explain why
effective field theories are necessary and useful.
Then four important cases are reviewed:
Symanzik's effective field theory of lattice spacing effects;
heavy-quark effective theory as a tool for controlling discretization
effects of heavy quarks;
chiral perturbation theory as a tool for reaching the chiral limit;
and a general field theory of hadrons for deriving finite volume
corrections.}

\vspace{1cm}
\tableofcontents
\newpage

\section{Introduction}

The idea to use lattice gauge theory to study quantum chromodynamics
(QCD) was introduced in 1974 in a seminal paper by Kenneth
Wilson.\cite{Wilson:1974sk}
It was an exciting time for the strong interactions:
it had become clear that quarks are partons,\cite{Bjorken:1969ja}
the Lagrangian for the gauge theory of quarks and gluons had just
been published,\cite{Fritzsch:1973pi}
and the discovery of asymptotic freedom was
new.\cite{Gross:1973id,Politzer:1973fx}
Soon afterwards an experiment at Brookhaven\cite{Aubert:1974js}
observed a new resonance, the~$J$, in proton-Beryllium collisions,
and an experiment at SLAC\cite{Augustin:1974xw} also observed a new
resonance, the~$\psi$, in $e^+e^-$ collisions.
This bound state of a charmed quark and its anti-quark, now called
the $J/\psi$, and its excitation spectrum has gone on to play a role
to similar to that of positronium in quantum electrodynamics.
Even today, the charmonium system plays an important
role in lattice~QCD.

The emergence of QCD as the theory of the strong interactions means
that high-energy scattering of partons and bound-state properties of
hadrons have a common explanation.
Owing to asymptotic freedom,\cite{Gross:1973id,Politzer:1973fx}
the gauge coupling in QCD becomes weaker at short distances, making
perturbative cross sections more accurate at higher energies.
The flip side, however, is that the coupling becomes stronger at long
distances: perturbation theory breaks down and the bound-state problem
in QCD is intrinsically non-perturbative.
Thus, most analysis of non-perturbative QCD has a general nature,
relying on, for example,
symmetries, analyticity, unitarity, and the renormalization group.
At its most successful, this kind of analysis yields quantitative
relationships between experimentally measurable quantities.
Perhaps the most striking example is to obtain parton densities from
deeply inelastic scattering, and then to use them to predict jet cross
sections in $p\bar{p}$ collisions.

Despite such successes, there is a need for general purpose tools
to calculate properties of the hadrons from the QCD Lagrangian, from
first principles.
One would like to see, quantitatively, that QCD can explain both jet
cross sections \emph{and} the proton mass, only by adjusting the QCD
coupling and the quark masses.
One would like to gain insight into the mechanisms of confinement,
of the type provided by detailed calculations.
Finally, to understand interactions of quarks at the shortest distances,
where new phenomena may be at play, it is usually necessary to have
calculations of certain hadronic matrix elements, with controlled,
comprehensible uncertainties.

Lattice gauge theory provides a mathematically well-defined framework
for non-perturbative QCD.
Before Wilson, Wegner\cite{Wegner:1971qt} had studied a version of
the Ising model, on a two-dimensional lattice, with a discrete gauge
symmetry.
Wilson realized how to implement the continuous SU(3) gauge symmetry
of QCD and other gauge theories and, more importantly, that lattice
field theory provides a non-perturbative definition of the functional
integral.
To make any sense of quantum field theory, an ultraviolet cutoff must
be introduced, and the key feature of lattice field theory is that it
introduces the cutoff at the outset.
With the lattice cutoff, it is possible to derive rigorously many
properties of field theory, as one can see, for example,
in the textbook of Glimm and Jaffe.\cite{GlimmJaffe}

\begin{figure}
	\centering
	\includegraphics[width=0.5\textwidth]{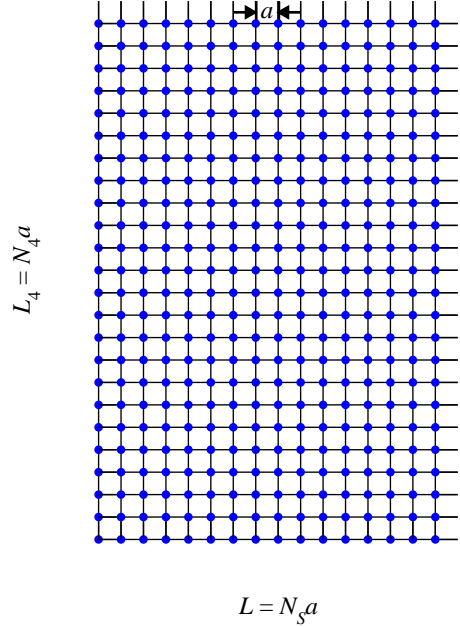}
	\caption[fig:lat]{Two-dimensional slice of a lattice with spacing $a$, 
	spatial size $L=N_Sa$, and temporal extent $L_4=N_4a$.}
	\label{fig:lat}
\end{figure}

The basic idea is to replace continuous spacetime with a discrete
lattice, usually hypercubic, as sketched in Fig.~\ref{fig:lat}.
The spacing between sites is usually denoted~$a$.
For simplicity we consider the spatial volume to have length $L=N_Sa$ 
on each side, and the temporal extent to be $L_4=N_4a$.
From a theoretical point of view, the lattice and finite volume provide 
gauge-invariant ultraviolet and infrared cutoffs, respectively.
Fermion fields $\psi(x)$ and $\bar{\psi}(x)$ live on sites~$x$.
Gauge fields live on links through the variables
\begin{equation}
	U_\mu(x)={\rm P}\exp\int_0^a ds\,A_\mu(x+se_\mu)
	\label{eq:Umx}
\end{equation}
where ${\rm P}$ denotes path ordering,
and $e_\mu$ is a unit vector in the $\mu$ direction.
In quantum field theory, information is obtained from correlation
functions, which have a functional integral representation.
In lattice QCD the correlation functions are
\begin{equation}
	\langle O_1\cdots O_n \rangle = \frac{1}{Z}
		\int \prod_{x,\mu} dU_\mu(x) \prod_x d\psi(x) d\bar{\psi}(x)
		\, O_1\cdots O_n \, e^{-S_{\rm QCD}}
	\label{eq:funcint}
\end{equation}
where $Z$ is defined so that $\langle 1\rangle=1$, and
$S_{\rm QCD}=-\sum_x{\cal L}_{\rm QCD}(x)$ is the (lattice) QCD action.
The $O_i$ are operators for creating the hadrons of interest and
also terms in the electroweak Hamiltonian.
With quarks on sites and gluons on links, it is possible to devise
lattice actions that respect gauge symmetry.
As in discrete approximations to partial differential equations,
derivatives in the Lagrangian are replaced with difference operators.
A~simple Lagrangian, introduced by
Wilson,\cite{Wilson:1974sk,Wilson:1975hf} is
\begin{eqnarray}
	{\cal L}_{\rm W} & = & {\cal L}_{\rm Wg} + {\cal L}_{\rm Wq}
	\label{eq:WilsonAction} \\
	{\cal L}_{\rm Wg} & = & \frac{1}{g_0^2a^4}\sum_{\mu\nu}\real\tr\left[
		U_\mu(x) U_\nu(x+ae_\mu) U^\dagger_\mu(x+ae_\nu) U^\dagger_\nu(x)
		-1\right]
	\label{eq:WilsonGaugeAction} \\
	{\cal L}_{\rm Wq} & = & - m_0 \bar{\psi}(x)\psi(x)
		- \sum_\mu \bar{\psi}(x)\left[
			P_{+\mu} {D_\mu^-}_{\text{lat}} -
			P_{-\mu} {D_\mu^+}_{\text{lat}}\right]\psi(x),
	\label{eq:WilsonQuarkAction}
\end{eqnarray}
where $g_0^2$ is the bare gauge coupling, and $m_0$ the bare quark mass.
(The literature often uses $\beta=6/g_0^2$ and $\kappa=(8+2m_0a)^{-1}$.)
In Eq.~(\ref{eq:WilsonQuarkAction})
\begin{eqnarray}
	P_{\pm\mu} & = & \case{1}{2}\left(1\pm\gamma_\mu\right), \\
	{D_\mu^+}_{\text{lat}}\psi(x) & = & a^{-1}\left[
		U_\mu(x)\psi(x+ae_\mu) - \psi(x) \right], \\
	{D_\mu^-}_{\text{lat}}\psi(x) & = & a^{-1}\left[
		\psi(x) - U^\dagger_\mu(x-ae_\mu)\psi(x-ae_\mu) \right].
\end{eqnarray}
The lattice clearly breaks spacetime symmetries, and we shall have to
confront that issue below.
But it preserves SU(3) gauge symmetry, and that is the most important
feature of Wilson's formulation of field theory.

It is a simple exercise to show that Wilson's Lagrangian reproduces 
the Yang-Mills Lagrangian in the naive continuum limit, 
$a\to 0$ with $g_0^2$ and $m_0$ fixed.
The bare couplings are, of course, unphysical, and the real continuum 
limit must be taken with physical quantities (e.g., hadron masses) 
held fixed.
Many discretizations have the correct naive continuum limit.
As long as the lattice Lagrangian is local,%
\footnote{Here ``local'' means that couplings for interactions of 
fields separated by a distance~$r$ fall off exponentially with $r$.}
they are all expected to share the same quantum continuum limit 
(with physical masses fixed).
The argument for such ``universality'' is based on the renormalization 
group, which indicates that physics depends only only the gauge 
couplings and quark masses.\cite{Wilson:1973jj}
These arguments have been tested experimentally in condensed matter 
systems, and although a rigorous proof has not been achieved,
there is a lot of mathematical evidence that lattice field theory 
defines quantum field theory.\cite{GlimmJaffe}

The study of QCD, which a theory of the natural world, is less concerned
with rigorous theorems than it is with practical results.
From this point of view, the breakthrough of the lattice formulation
is that Eq.~(\ref{eq:funcint}) turns quantum field theory into a
mathematically well-defined problem in statistical mechanics.
Condensed matter theorists and mathematical physicists have devised a
variety of methods for tackling such problems.
In addition to weak-coupling perturbation theory, the toolkit
includes non-perturbative versions of the renormalization group,
strong coupling expansions, and numerical integration of the functional
integral by Monte Carlo methods.
The first two were pursued with great vigor in the decade following
Wilson's original paper.
The strong coupling limit is especially appealing, because confinement
emerges immediately, cf.\ Sec.~\ref{sec:ten}.
The last is the most widely used today, especially for problems 
motivated by particle physics, nuclear physics, or astrophysics.
Indeed, when most physicists speak of lattice QCD, 
they mean numerical lattice QCD.

It is difficult to decide what aspects of lattice QCD to cover in a
single chapter of a handbook of all~QCD.
There are several textbooks on lattice gauge
theory,\cite{Creutz:mg,Rothe:kp,Montvay:cy}
with emphasis on lattice QCD.
These books, as well as many review articles%
\cite{Kogut:1983ds,Hasenfratz:1985pd,Kronfeld:1993jf,Aoki:1999ue,%
Kenway:2000pp}
and summer school lecture series,%
\cite{DeGrand:1990ss,Luscher:1988sd,Kronfeld:1992hw,Sharpe:1994dc,%
Gupta:1997nd,Luscher:1998pe,Munster:2000ez,DeGrand:2000qg}
cover the foundations well.
Nevertheless, many colleagues---theorists whose research is in continuum
QCD and experimenters whose measurements need non-perturbative QCD
to be interpreted---tell me that papers on lattice calculations are
impenetrable.

One complaint is that the explanations of numerical techniques are
written in an unfamiliar jargon, which is unfortunate but hard to
rectify here.
A~more serious complaint surrounds the uncertainties that arise in
moving the idealized problem of mathematical physics to a practical
problem of computational physics.
Many non-experts know that these arise from Monte Carlo statistics,
non-zero lattice spacing, finite spacetime volume, and unphysical values
(in the computer) of the quark masses.
But, nevertheless, the methods to deal with them are (evidently)
not transparent.
This is a shame.
Everyone has a feel for statistical errors, even without knowing how 
to write a Monte Carlo program to evaluate the right-hand side of 
Eq.~(\ref{eq:funcint}).
Similarly, the other uncertainties are controlled and understood with
effective field theories, so a basis for a common language should be
possible.

Two examples illustrate why a common understanding of the uncertainties
is needed.
The first comes from flavor physics, where a wide variety of $B$,
$D$, and $K$ decays are studied, to test whether the standard
Cabibbo-Kobayashi-Maskawa (CKM) mechanism adequately explains
flavor and $CP$ violation.
The quark-level CKM interpretation of many of these processes is 
obscured by the uncertainty in hadronic matrix elements, of the type 
$\bracket{f}{{\cal H}}{H}$, where $H$ is a strange, charmed, or 
$b$-flavored hadron, and ${\cal H}$ is a term in the electroweak 
Hamiltonian.
${\cal H}$ arises from integrating out $W$, $Z$, $t$, and (possibly)
other more massive particles.
Experimental measurements are, or soon will be, very precise.
Without reliable theoretical calculations, including a transparent
error analysis, it will be much more difficult tell whether ${\cal H}$
is solely electroweak in origin or, perhaps, has non-standard
contributions.
Lattice calculations are most straightforward when the final state~$f$
has leptons and either one hadron or none.
\footnote{When there are two hadrons in the final state, 
there are additional complications, cf.\ Sec.~\ref{subsec:V}.}
Such matrix elements are needed for leptonic, radiative, and 
semi-leptonic decays, as well as for neutral meson mixing.

A second example, though less often mentioned, is the search for new 
particles at the Tevatron and, later, at the~LHC.
If a new particle carries color, then its decays always contain jets, 
and observation depends on the extra jet production standing out 
against normal QCD jet production.
The QCD uncertainty has, in the past, been estimated by comparing
various fits to the parton densities.
A~closer look,\cite{Giele:2001mr} however, reveals that the parton
densities are well-constrained over a limited range of~$x$
($x$:~fraction of the proton momentum taken by a parton).
Uncontrolled uncertainties in predictions of cross sections arise 
because there are only meager constraints on the parton densities 
outside this range.
In lattice gauge theory, however, it is possible to calculate the 
moments of the parton densities: they are related via the 
operator-product expansion to matrix elements of local operators.
By the time the LHC experiments run, it should be possible to obtain
the first few moments with an uncertainty that is not only small,
but also well justified.\cite{Gockeler:1995wg,Guagnelli:2000sf,%
Dolgov:2002mn,Jansen:2000xm}
The direct calculation of moments and the direct measurement over a
finite interval in $x$ will provide complementary information, making
possible signals of new physics more persuasive.

Suppose the comparison of theory and experiment, in either flavor 
physics or highest-energy collisions, leads to a hypothesis
of new, non-standard phenomena.
Then the stakes become very high, and the reliability of error bars 
becomes the central concern.
It seems, therefore, worthwhile to discuss how to control and estimate 
uncertainties in numerical calculations, and this issue is the central 
theme of this chapter.
The origin of statistical errors in numerical lattice QCD is reviewed 
in Sec.~\ref{sec:mc}.
Then the origin of systematic uncertainties is reviewed in 
Sec.~\ref{sec:why}, motivating the main tools for controlling them.
In most cases the tool is an effective field theory, which allows us
to control the extrapolation of artificial, numerical data to the real
world, provided the data start ``near'' enough.
Most particle physicists are familiar with the logic and utility of 
effective field theories, and know how to judge their range of 
validity.
Thus, effective field theories should also provide a common language 
for experts and non-experts to discuss the error bars, without 
requiring the non-experts to repeat all the steps of the numerical 
analysis.
The effective field theories needed to analyze the controllable 
uncertainties are discussed in Secs.~\ref{sec:sym}--\ref{sec:volume}.

An uncontrolled systematic effect of many lattice calculations has 
historically been the quenched approximation.
It is reviewed in Sec.~\ref{sec:mc}.
It is hard to estimate the associated error, and only in isolated 
cases can one argue that it is a subdominant error, let alone that it is
under control.

Section~\ref{sec:ten} contains a ``top ten'' list of the trends 
and developments in lattice QCD, which warrant appreciation.
These have been chosen for their broad interest, and because they should
influence one's thinking about QCD.

\section{Overview of Numerical Techniques}
\label{sec:mc}

This section gives a brief overview the numerical methods.
These are covered in more detail in some of the texts, reviews, and 
summer schools cited above, as well as in a set of lecture notes aimed 
at experimenters.\cite{DiPierro:2000nt}

The foremost issue is that there are very many variables.
Continuum field theory has uncountably many degrees of freedom.
Field theory on an infinite lattice still has an infinite number of
degrees of freedom, but the infinity is now countable, {\em i.e.}, it 
is as infinite as the integers.
This makes the products over $x$ in Eq.~(\ref{eq:funcint})
well-defined.
For a computer (with finite memory), the number of degrees of freedom 
must be kept finite.
To do so, one must also introduce a finite spacetime volume.
This may seem alarming, but what one has done is simply to introduce an
ultraviolet cutoff (the lattice) and an infrared cutoff (the finite
volume).
All calculations in QCD, except trivial ones, require an ultraviolet 
cutoff, and many require an infrared cutoff, although physical
predictions are cutoff independent.
In a sense, removal of the cutoffs is the subject of 
Secs.~\ref{sec:sym} and~\ref{sec:volume}.

Even with a finite lattice, the number of integration variables is
large.
For QCD on a $N_S^3\times N_4$ lattice (cf.\ Fig.~\ref{fig:lat}) there 
are $(4\times8)N_S^3N_4$ variables for gluons
and $(4\times3)N_S^3N_4$ for quarks.
If one only demands a volume a few times the size of a hadron and also
several grid points within a hadron's diameter, one already requires
at least, say, 10 points along each direction.
In four-dimensional spacetime this leads to $\sim 32\times 10^4$ gluonic
variables.

With so many variables, the only feasible methods are based on Monte
Carlo integration.
The basic idea of Monte Carlo integration is simple: generate an
ensemble of random variables and approximate the integrals in
Eq.~(\ref{eq:funcint}) by ensemble averages.
Thus, calling all variables~$\phi$,
\begin{equation}
	\langle O_1\cdots O_n\rangle = \frac{1}{Z} \sum_{z=1}^{N_Z}
		w(\phi^{(z)}) O_1(\phi^{(z)})\cdots O_n(\phi^{(z)})
	\label{eq:ensemble}
\end{equation}
with weights~$w$ to be specified below, 
and $Z$ defined so that $\langle1\rangle=1$.

Quarks pose special problems, principally because, to implement Fermi
statistics, fermionic variables are Grassmann numbers.
In all cases of interest, the quark action can be written
\begin{equation}
	S_{\rm q} = -\sum_x{\cal L}_{\rm q}(x) = \sum_{\alpha\beta}
		\bar{\psi}_\alpha M_{\alpha\beta} \psi_\beta,
\end{equation}
where $\alpha$ and $\beta$ are multi-indices for (discrete) spacetime,
spin and internal quantum numbers.
The matrix $M_{\alpha\beta}$ is some discretization of the Dirac
operator $\kern+0.1em /\kern-0.65em D+m$, such as
Eq.~(\ref{eq:WilsonQuarkAction}).
Note that it depends on the gauge field, but one may integrate over the
gauge fields after integrating over the quark fields.
Then, because the quark action is a quadratic form, the integral can be
carried out exactly:
\begin{equation}
	\int \prod_{\alpha\beta} d\bar{\psi}_\alpha d\psi_\beta\,
		e^{- \bar{\psi} M \psi} = \det M .
	\label{eq:quarkdet}
\end{equation}
Similarly, products $\psi_\alpha\bar{\psi}_\beta$ in the integrand 
$O_1\cdots O_n$ are replaced with quark 
propagators~$[M^{-1}]_{\alpha\beta}$ using the familiar rules of Wick 
contraction.
The computation of $M^{-1}$ is demanding, and the computation of $\det M$
(or, more precisely, changes in $\det M$ as the gauge field is changed)
is very demanding.

With the quarks integrated analytically, it is the gluons that are
subject to the Monte Carlo method.
The factor weighting the integrals is now $\det M e^{-S_{\rm g}}$, 
where $S_{\rm g}$ is the gluons' action.
Both $\det M$ and $e^{-S_{\rm g}}$ are the exponential of a number 
that scales with the spacetime volume.
In Minkowski spacetime the exponent is an imaginary number, so
there are wild fluctuations for moderate changes in the gauge field.
On the other hand, in Euclidean spacetime, with an imaginary time
variable, $S_{\rm g}$ is real.
In that case (and assuming $\det M$ is non-negative) one can devise a 
Monte Carlo with \emph{importance sampling}, which means that the 
random number generator creates gauge fields weighted according to 
$\det M e^{-S_{\rm g}}$.
With importance sampling the weights on the right-hand side of 
Eq.~(\ref{eq:ensemble}) are independent of the fields, 
so one can set~$w=1$.
Because importance sampling is necessary to make lattice QCD 
numerically tractable, all numerical work is done in Euclidean 
spacetime.

Importance sampling works well if $\det M$ is positive.
For pairs of equal-mass quarks, this is easy to achieve.
With the Wilson action, Eq.~(\ref{eq:WilsonQuarkAction}),
$\gamma_5M\gamma_5=M^\dagger$.
Since this is a similarity transformation, $M$ and $M^\dagger$ have 
the same physical content.
With $M$ for one flavor and $M^\dagger$ for the other (of same mass),
the fermion determinant is $\det(M^\dagger M)$, which is obviously 
non-negative.
The same argument holds for Neuberger's
discretization,\cite{Neuberger:1997fp}
which is computationally more demanding, but has better chiral 
symmetry (cf.\ Sec.~\ref{sec:chiral}).
For the Kogut-Susskind quark action,\cite{Susskind:1977ks} the matrix 
$M_{\rm KS}$ is non-negative, but each Kogut-Susskind field creates 4 
fermion species in the continuum limit.

Thus, most calculations of $\det M$ are for 2 or 4 flavors.
The physically desirable situation with three flavors, with the
strange quark's mass different from that of two lighter quarks, is 
difficult to achieve.
One way is to cope with occasionally negative 
weights.\cite{Aoki:2001pt}
Some algorithms for generating the gauge fields set up a guided random 
walk with a finite step size~$\epsilon$.%
\cite{Duane:1985ym,Ukawa:1985hr,Batrouni:1985jn}
In a widely used scheme for introducing the
fermions,\cite{Batrouni:1985jn} they can generate weights such as
$(\det M^\dagger M)^{1/2}$ or $(\det M_{\rm KS})^{1/4}$,
which formally give a single flavor.
There is some evidence that there may be subtleties associated with 
non-zero~$\epsilon$ in large systems with small quark 
masses.\cite{Sexton:2002pi}
These potential problems, and also the physical interpretation of the 
fractional powers, could be monitored by looking at the pattern of 
spontaneously broken chiral symmetry, cf.\ Sec.~\ref{sec:chiral}.

The choice of imaginary time has an important practical advantage.
Consider the two-point correlation function
\begin{equation}
	C_2(t) = \bra{0} \Phi_H(t) \Phi_H^\dagger(0) \ket{0},
	\label{eq:C2}
\end{equation}
where $\Phi_H$ is an operator with the quantum numbers of the hadron 
of interest,~$H$.
For simplicity, assume $t>0$ and take the total spatial momentum to 
vanish.
Inserting a complete set of eigenstates of the Hamiltonian between 
$\Phi_H$ and $\Phi_H^\dagger$,
\begin{equation}
	C_2(t) = \sum_n \bra{0} \Phi_H \ket{H_n}
		\bra{H_n} \Phi_H^\dagger \ket{0} e^{im_{H_n}t},
\end{equation}
where $m_{H_n}$ is the mass of $\ket{H_n}$, the $n$th radial excitation
with $H$'s quantum numbers.
For real $t$ it would be difficult to disentangle all these
contributions.
If, however, $t=ix_4$, with $x_4$ real and positive, then one has a sum
of damped exponentials.
For large $x_4$ the lowest-lying state dominates and
\begin{equation}
	C_2(x_4) = |\bra{0} \Phi_H \ket{H}|^2 e^{-m_Hx_4} + \cdots,
	\label{eq:C2exp}
\end{equation}
where $\ket{H}$ is the lowest-lying state and $m_H$ its mass.
The omitted terms are exponentially suppressed.
It is straightforward to test when the first term dominates a
numerically computed correlation function, and then fit the exponential
form to obtain the mass.

This technique for isolating the lowest-lying state is also essential
for obtaining hadronic matrix elements.
For a transition from one hadron~$H$ to another~$H'$, 
one must compute the matrix element $\bra{H'}Q\ket{H}$, where
$Q$ is the operator inducing the transition.
In flavor phenomenology, $Q$ is a term in the electroweak Hamiltonian;
for moments of parton densities, $Q$ is a local operator appearing in 
the operator product expansion of two currents.
One uses a three-point correlation function
\begin{equation}
	C_{H'QH}(x_4,y_4) = \bra{0} \Phi_{H'}(x_4+y_4) Q(y_4) 
		\Phi^\dagger_H(0) \ket{0},
\end{equation}
where only the Euclidean times of the operators have been written out.
Inserting complete sets of states and taking $x_4$ and $y_4$ large
enough,
\begin{equation}
	C_{H'QH}(x_4,y_4) = \bra{0} \Phi_{H'} \ket{H'} \bra{H'} Q \ket{H}
		\bra{H} \Phi^\dagger_H\ket{0} e^{-m_{H'}x_4-m_Hy_4}.
	\label{eq:CQexp}
\end{equation}
The amplitudes $\bra{0}\Phi_{H'}\ket{H'}$ and 
$\bra{H}\Phi^\dagger_B\ket{0}$ and the masses $m_{H'}$ and $m_H$ 
are obtained from two-point correlation functions $C_2$,
leaving $\bra{H'}Q\ket{H}$ to be determined from~$C_{H'QH}$.
To compute amplitudes for a transition from $H$ to the vacuum (as in a 
leptonic decay), one can simply replace $\Phi_H$ in $C_2$ with the 
charged current.

Equations~(\ref{eq:C2})--(\ref{eq:CQexp}) assumed $t>0$,
$x_4+y_4>y_4>0$, {\em and} $L_4\to\infty$.
With finite~$L_4$ other time orderings lead to terms with
$e^{-m(L_4-t)}$.
They are straightforward to derive and to incorporate into fits.
Thus, these details do not alter the basic paradigm for computing 
masses and matrix elements.

These methods are conceptually clean and technically feasible for 
calculating masses and hadronic matrix elements with at most one 
hadron in the final state.
The procedure for computing correlation functions is as follows.
First generate an ensemble of lattice gluon fields with the appropriate
weight.
Next form the desired product $O_1\cdots O_n$, with quark variables
exactly integrated out to form propagators~$M^{-1}$.
Then take the average over the ensemble.
Finally, fit the Euclidean time dependence of 
Eqs.~(\ref{eq:C2exp}) and~(\ref{eq:CQexp}).
With two hadrons in the final state, correlation functions can be 
obtained in more or less the same way, but the interpretation of the 
energies and amplitudes is more complicated, as discussed in 
Sec.~\ref{subsec:V}.

Within the same ensemble, there are correlations in the statistical 
fluctuations of the quantities calculated.
Methods, such as bootstrap and jackknife, that propagate correlations 
through the analysis are well understood and widely used.
So, these days, statistical errors rarely lead to controversy.
As discussed in Sec.~\ref{sec:why}, it is not practical to carry out 
the Monte Carlo calculations at very small lattice spacings or at very 
small quark masses.
To gain control over these effects (using effective field theory to 
guide extrapolations to the physical limit) requires small statistical 
errors on the raw output of the Monte Carlo calculation.


As mentioned above, the computation of the factor $\det M$ in 
Eq.~(\ref{eq:quarkdet}) is very demanding.
The determinant generates sea quarks inside a hadron.
It is thus tempting to replace $\det M$ with~$1$ \emph{and} 
compensate the corresponding omission of the sea quarks with shifts 
in the bare couplings.
This approximation is most often called the quenched 
approximation.\cite{Marinari:1981qf}
A~more vivid name is the valence approximation,\cite{Weingarten:1981jy}
which stresses that the valence quarks (and gluons) in hadrons are
treated fully, and the sea quarks merely modeled.
The idea is analogous to a dielectric approximation in electromagnetism,
and it fails under similar circumstances.
In particular, if one is interested in comparing two quantities that 
are sensitive to different energy scales, one cannot expect 
the same dielectric shift to suffice.

It is not easy to estimate quantitatively the effect of quenching.
The quenched approximation can be cast as the first term in a
convergent expansion,\cite{Sexton:1997ud} providing a method to 
compute the shifts in the couplings, and further corrections.
The computed shift agrees with the empirical one, but it is about as 
difficult to compute the next term as to restore the fermion 
determinant.
For some quantities one can estimate the short-distance contribution 
to the quenching shift.
Examples include the strong coupling~$\alpha_s$\cite{El-Khadra:1992vn} 
the quark masses\cite{Mackenzie:1994zw,Davies:1994zw,Gough:1996zw},
and ${\cal F}(1)$, which is a form factor needed to determine the CKM
matrix element~$V_{cb}$.\cite{Hashimoto:2001ds}
It is the long-distance part which is harder to fathom.

The quenched approximation is going away.
In heavy quark physics the {CP-PACS}\cite{AliKhan:2000eg,AliKhan:2001jg}
and MILC\cite{Bernard:2000nv} collaborations have unquenched
calculations of the heavy-light decay constants $f_B$, $f_{B_s}$,
$f_D$, and~$f_{D_s}$.
Both groups have results at several lattice spacings, so they can 
study the $a$~dependence.
Their results are about 10--15\% higher than the most mature estimates 
from the quenched approximation.
In addition, the Rome group has an unquenched calculation of the $b$ 
quark mass, which agrees well with their quenched 
calculation.\cite{Gimenez:2000cj}
There are also unquenched calculations of moments of parton
densities.\cite{Dolgov:2002mn}

\section{Why Effective Field Theories?}
\label{sec:why}

In this section we discuss why it is necessary, as a practical matter, 
to consider effective field theories.
The first clue is that the physical problem has many scales, so one
should think of renormalization-group strategies to tackle them.
Effective field theory is one of the most powerful such tools.
The central idea is to introduce a separation scale
(in energy units)~$\mu$.
Effects from distances shorter than $\mu^{-1}$ are lumped into the 
couplings (or short-distance coefficients) of the effective field 
theory, whereas effects from distances longer than $\mu^{-1}$ are 
described by operators in the effective field theory.
The degrees of freedom in the effective field theory are those required
to reproduce the singularities of thresholds,\cite{Coleman:1965le}
and so on, of the underlying theory, when only processes of energy
$E<\mu$ are considered.
By demanding that physical results do not depend on $\mu$, and by
matching the effective to the underlying theory, one can avoid
over- or under-counting contributions at the interface of ``long'' and
``short.''

In QCD, the energy scale characteristic of non-perturbative gluonic 
effects is called~$\Lambda$, and it lies in a range from
the asymptotic freedom parameter
$\Lambda_{\overline{\rm MS}}\approx 250$~MeV
through the $\rho$~meson mass $m_\rho=770$~MeV to 
the scale of chiral symmetry breaking
$m_K^2/m_s\approx 2500$~MeV. 
For concreteness, we shall think of $\Lambda\approx750$~MeV, allowing
leeway of a factor of three where appropriate.
QCD also has quarks, whose masses range widely.
Light quarks are those with $m_q<\Lambda$ or even $m_q\ll\Lambda$;
the mass of the strange quark is about 100~MeV---seven or eight times 
smaller than~$\Lambda$, and the masses of the up and down quarks are 
about 25~times smaller still.
Heavy quarks are those with $m_Q>\Lambda$ or even $m_Q\gg\Lambda$; 
the mass of the bottom quark is 4--5~GeV---roughly six times larger 
than~$\Lambda$, and the mass of the top quark is about 40~times larger 
still.
The mass of the charmed quark is slightly larger than~$\Lambda$, 
$m_c\approx1.3$~GeV.
Whether, or under what circumstances, the charmed quark can be 
treated as heavy is an open question.
Systems with valence charmed quarks seem to enjoy some of the
simplifications of analogous systems with valence $b$ quarks.
On the other hand, a sea of $c\bar{c}$ pairs could play a role in the 
high-momentum tail of hadronic wave functions, where not much is known.

In numerical calculations, one also has cutoffs.
The lattice has a non-zero spacing~$a$, corresponding to a
ultraviolet cutoff $\sim\pi/a$.
The spacetime volume has a finite size~$L^3L_4$, corresponding to a
infrared cutoff $(L^3L_4)^{1/4}$.
With arbitrarily large computer memory and processing power, 
one could imagine taking these scales in the hierarchy
\begin{equation}
	L^{-1} \ll m_q \ll \Lambda \ll m_Q \ll a^{-1},
	\label{eq:impractical}
\end{equation}
with the quark masses adjusted to their physical values.
(Because quarks are confined, the adjustment is made by tuning one
hadron mass for each flavor of quark.)
In that case one would only need to know a little about cutoff effects:
just enough to be confident that the cutoffs introduce only small
deviations from the limits $a\to 0$ and $L\to\infty$.

In practice, however, this idealized situation cannot be achieved.
Finite computer memory and processor power limit both
$a$ and~$L$.
To get a feel for the tradeoffs, one needs only a few simple scaling
laws for the computer algorithms.
First, the amount of memory needed grows as
\begin{equation}
	{\tt memory} \propto N_S^3N_4 = L^3L_4/a^4.
	\label{eq:memory}
\end{equation}
Increasing either the physical volume $L^3$ or decreasing the lattice
spacing~$a$ puts great demands on the memory.
The large exponents in Eq.~(\ref{eq:memory}) are unavoidable, because
they stem from the fact that we live in $3+1$ dimensions.
Second, the amount of CPU time needed to create statistically
independent lattice gauge fields (with $L$ fixed) grows as
\begin{equation}
	\tau_g \propto a^{-(4+z)}.
	\label{eq:gaugeslow}
\end{equation}
The~4 in the exponent arises because the number of variables to process
grows as $a^{-4}$.
In addition, the update algorithms slow down as $a\to 0$, because
they update in a region of size $a$, but must propagate these changes
over physical regions of size $\Lambda^{-1}$ to get a statistically
independent gauge field.
Thus, the exponent $z>0$, and available algorithms have $z$
around~1 or~2.\cite{Kronfeld:1992jf}

In addition to creating gauge fields, one must compute quark propagators
in the background gauge field.
These propagators are needed for the valence quarks in any hadron.
The numerical problem is to solve
\begin{equation}
	MG = S
	\label{eq:MG=S}
\end{equation}
for $G$, given some source~$S$, where $M$ is the discretized Dirac
operator.
$M$, $G$, and $S$ are ${\Bbb N}\times{\Bbb N}$ matrices, where
${\Bbb N}\propto N_S^3N_4$ is number of quark degrees of freedom.
The matrix $M$ is sparse (meaning that most of the entries vanish),
because it is practical only to put the most local interactions into
the lattice Lagrangian.
The CPU time needed to solve for $G$, even with the best algorithms, is
\begin{equation}
	\tau_q \propto (\lambda_{\rm max}/\lambda_{\rm min})^p
		\sim \min\left\{1/(m_qa)^p, (L/a)^p\right\}.
	\label{eq:mqslow}
\end{equation}
where $\lambda_{\rm max}$ and $\lambda_{\rm min}$ are the largest and
smallest eigenvalues of $M$, and $m_q$ is the quark mass.
The exponent $p$ depends on the algorithm and is typically 1~or~2.
Equation~(\ref{eq:MG=S}) is also needed in algorithms to incorporate 
sea quarks correctly,\cite{Duane:1985ym,Ukawa:1985hr,Batrouni:1985jn,%
Gottlieb:1987mq,Duane:1987de,Luscher:1993xx}
where the effective exponent~$p$ is 2~or~3.
At fixed $a$ it is, therefore, costly to reduce $m_q$.
For suitable boundary conditions, the volume term in
Eq.~(\ref{eq:mqslow}) can take over when $m_q<L^{-1}$, but this
regime has significant finite size effects, so it is not suited to
general-purpose hadron phenomenology.
In addition to difficulties with finite-size effects and with the
chiral slowdown of the algorithms, statistical uncertainties increase 
as the quark mass decreases.
The bottom line is that it is not practical to run the computer at
masses as small as those of the up and down quarks in nature.

\begin{figure}
	\centering
	\includegraphics[width=\textwidth]{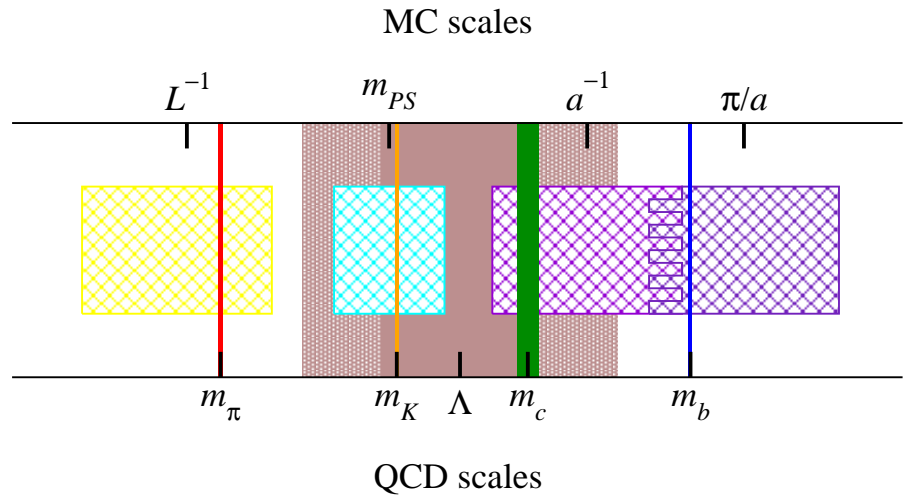}
	\caption[fig:scales]{Scales in QCD, and in numerical lattice
	calculations.
	Physical scales of QCD are labeled on the bottom, 
	and are indicated by solid colors.
	Scales from practical Monte Carlo calculations are labeled at the 
	top, and indicated by cross-hatched patches.}
	\label{fig:scales}
\end{figure}

In typical calculations, these days, $N_S=16$--32,
with $N_4$ the same or a few times larger.
To balance the infrared and ultraviolet cutoffs, one ends up with 
$a^{-1}\sim1$--4~GeV (so $\pi/a\sim3$--12~GeV) and $L\sim1$--4~fm.
In typical calculations the ``light'' quarks have mass in the range
0.2--$0.4<m_q/m_s<1.2$.
Thus, in practice, the idealized hierarchy~(\ref{eq:impractical})
becomes
\begin{equation}
	L^{-1} < m_q < \Lambda \ll m_b \sim a^{-1},
	\label{eq:fermi-practical}
\end{equation}
which is sketched in Fig.~\ref{fig:scales}.
Although the various scales are not as well separated as in the
idealized hierarchy~(\ref{eq:impractical}), there is still some
separation.
Thus, one has a chance of using effective field theory to go from
sequences of calculations roughly in the
hierarchy~(\ref{eq:fermi-practical}) to continuum, infinite-volume QCD
with the quark mass of the real world.

The large masses of the bottom and charmed quarks make the numerical
solution of Eq.~(\ref{eq:MG=S}) relatively easy, but heavy-quark
discretization effects become a difficult problem, and a topic of
some debate.
Instead of studying $m_Q\sim a^{-1}$ directly, some groups set $m_Q<m_b$
(and, indeed, $m_Q\lesssim m_c$), leading to the hierarchy
\begin{equation}
	L^{-1} < m_q \lesssim \Lambda  <  m_Q < a^{-1}.
	\label{eq:euro-practical}
\end{equation}
Another approach for heavy quarks is the static approximation,
$m_Q\to\infty$.
Methods for heavy quarks are discussed further in Sec.~\ref{sec:hqet}.

In summary, practical limitations of computers constrain the
lattice spacing, the quark masses, and the box size to the
hierarchy~(\ref{eq:fermi-practical}) or~(\ref{eq:euro-practical})
instead of the idealized hierarchy~(\ref{eq:impractical}).
Nevertheless, these parameters can all be varied over a certain range,
providing numerical lattice QCD with one of its most important
strengths.
The paradigm is as follows:
the computer generates numerical data, varying each of $a$, $m_Q$,
$m_q$, and $L$.
These data must then be analyzed to extract~QCD, at least as
long as the data start close enough to the real world.
With sound theoretical guides, this paradigm is practical, and allows 
propagation of errors.\cite{Lepage:2001ym}
In each case, the guide comes from an effective field theory.
Moreover, one can test the functional form anticipated
from an effective field theory and then---assuming the test succeeds---%
the extrapolation to the physical limit is justified.
Indeed, this paradigm uses limited computer power much more 
effectively than when relying on brute force alone.

For reference, the most important effective field theories are listed 
in Table~\ref{tab:eft}.
\begin{table}
	\caption[tab:eft]{Effective field theories used in controlling and
	quantifying systematic uncertainties of lattice calculations.
	(NRQCD is used for bound states of a heavy quark and a heavy
	anti-quark, called quarkonium, where the small parameter is the
	relative velocity~$\nu$ between the constituents.)}
	\label{tab:eft}
	\begin{center}
	\begin{tabular}{clc}
	\hline\hline
		scale & \multicolumn{1}{c}{EFT} & small parameter \\
	\hline
		 $a$  & Symanzik effective field theory & $\Lambda a$ \\
		$m_Q$ & heavy-quark effective theory (HQET) & $\Lambda/2m_Q$ \\
		      & non-relativistic QCD (NRQCD) & $\nu$ \\
		$m_q$ & chiral perturbation theory ($\chi$PT) & 
				$(m_K/4\pi f_\pi)^2\sim m_s/\Lambda$ \\
		 $L$  & L\"uscher effective field theory & 
		 		$e^{-m_\pi L}$, $(L\Lambda)^{-1}$ \\
	\hline\hline
	\end{tabular}
	\end{center}
\end{table}
Quark mass dependence relies on methods that are also used in 
continuum QCD: chiral perturbation theory ($\chi$PT) for light hadrons,
and heavy-quark effective theory (HQET) and non-relativistic QCD (NRQCD) 
for heavy quark systems.
A~glance at Fig.~\ref{fig:scales} shows that in most cases the scales
are not very far apart.
The leading term in each effective theory may not be enough to give a 
good guide to the extrapolation, but it is possible to work out 
higher-order expressions analytically and use them.

In some cases it makes sense to combine two or more of the tools.
As mentioned above, on typical lattices the $b$ quark satisfies
$m_ba\sim1$, so an effective theory treating both $a$ and $m_b^{-1}$ as
short distance should be useful.
This idea is developed in Sec.~\ref{sec:hqet}.
Most formulations of lattice fermions explicitly break some of the
(chiral) flavor symmetries of QCD.
To study the interplay of the continuum limit and spontaneous chiral
symmetry breaking, one can modify the chiral Lagrangian to parametrize 
the lattice's explicit chiral symmetry breaking.
We shall return to this point in Sec.~\ref{sec:chiral}.
Finally, when masses of pseudo-Goldstone bosons
($\pi$, $K$, $\eta$ in nature)
become small compared to the volume, it becomes necessary to study their
finite-size effects with chiral perturbation theory.
This topic is mentioned briefly in Sec.~\ref{sec:volume}.

\section{Lattice Spacing Effects: Symanzik Effective Field Theory}
\label{sec:sym}

The most obvious difference between lattice gauge theory and continuum
QCD is the non-zero lattice spacing.
It is often (correctly) said that lattice field theories {\em define}
quantum field theories.
But the definition entails taking a sequence of lattice theories, with
varying lattice spacing, and taking the limit $a\to 0$ with physical
quantities held fixed.\cite{Wilson:1973jj}
To interpret computer calculations at non-zero~$a$, however, what one
really needs is a description of cutoff effects.
This section discusses such a description, based on an effective field 
theory invented by Symanzik.\cite{Symanzik:1979ph,Symanzik:1983dc}
It provides simple semi-quantitative estimates of lattice spacing 
effects.
More interestingly, it provides strategies for eliminating them, both
by parametrically reducing their size, and by giving a framework for
combining results from several lattice spacings.

One can get some feeling for lattice spacing effects by looking at
the quark-gluon vertex.
With Wilson's action for lattice fermions, 
Eq.~(\ref{eq:WilsonQuarkAction}), the vertex is
\begin{eqnarray}
	\Gamma_\mu(p,p')
		&\!=\!& -g_0t^a\left\{ \gamma_\mu\cos[\case{1}{2}(p+p')_\mu a] -
			i \sin[\case{1}{2}(p+p')_\mu a] \right\} \nonumber \\
	 	&\!=\!& -g_0t^a\left\{ \gamma_\mu -
			\case{i}{2} a(p+p')_\mu + O(a^2) \right\}.
	\label{eq:WilsonVTX}
\end{eqnarray}
In a hadron at rest, the typical momenta are of order~$\Lambda$.
So, if $\Lambda a$ is around 0.2, the quark-gluon interaction deviates
by about 20\% from the continuum.

Recall, however, that the lattice action is not unique.
In 1985, Sheikh\-oleslami and Wohlert\cite{Sheikholeslami:1985ij} 
suggested adding another interaction to the Wilson action, 
namely a lattice approximant to $i\bar{q}\sigma\cdot Fq$, 
with coefficient~$c_{\text{SW}}/4$.
Then the quark-gluon vertex becomes
\begin{eqnarray}
	\Gamma_\mu(p,p')
		&=& -g_0t^a\left\{ \gamma_\mu\cos[\case{1}{2}(p+p')_\mu a] -
			i \sin[\case{1}{2}(p+p')_\mu a]
		\right. \nonumber \\ & & \left. +
			\case{1}{2}c_{\text{SW}}\sigma_{\mu\nu}
			\cos[\case{1}{2}k_\mu a]\sin[k_\nu a] \right\} \nonumber \\
	 	&=& -g_0t^a\left\{ \gamma_\mu -
			\case{i}{2} a\left[(p+p')_\mu + c_{\text{SW}}
			  i\sigma_{\mu\nu}k^\nu\right] + O(a^2) \right\},
	\label{eq:SWVTX}
\end{eqnarray}
where $k=p'-p$, $\sigma_{\mu\nu}=i[\gamma_\mu,\gamma_\nu]/2$.
On the mass shell, an easy application of the Gordon identity
shows that the two sets of $O(a)$ terms cancel if $c_{\text{SW}}=1$.

The analysis of Eqs.~(\ref{eq:WilsonVTX}) and~(\ref{eq:SWVTX}) is
essentially classical.
For quantum field theory, one would rather have a formalism that
allows for renormalization, and also gives a concept of ``on shell''
that holds for hadrons.
A few years after Wilson's paper, 
Symanzik\cite{Symanzik:1979ph,Symanzik:1983dc} introduced a local 
effective Lagrangian for analyzing discretization effects.
This work grew out of his earlier studies cutoff dependence in field 
theories, a line of thinking that had led to the Callan-Symanzik 
equation.\cite{Symanzik:1970rt,Callan:1970yg}
The idea is that lattice gauge theory, at given~$a$, can be described
by a local effective Lagrangian.
Short-distance effects, in particular discretization effects, are
lumped into short-distance coefficients, whereas long-distance physics
is described by local effective operators.
The idea is analogous to the usage of effective field theories to 
parametrize short-distance phenomena whose microscopic dynamics is 
unknown.

For the Lagrangian of any lattice field theory, Symanzik says to 
write\cite{Symanzik:1979ph}
\begin{equation}
	{\cal L}_{\text{lat}} \doteq {\cal L}_{\text{Sym}},
	\label{eq:lat=Sym}
\end{equation}
where the symbol $\doteq$ can be read
``has the same on-shell matrix elements as.''
The left-hand side of Eq.~(\ref{eq:lat=Sym}) is the lattice field 
theory inside the computer, whose output must be analyzed to obtain a 
continuum result.
The lattice action can be complicated, with many free parameters.
For lattice QCD with Wilson fermions
\begin{equation}
	{\cal L}_{\text{lat}} = {\cal L}_{\text{W}} +
		\sum_O a^{\dim O-4} c_O\, O_{\text{lat}},
	\label{eq:Llat}
\end{equation}
where ${\cal L}_{\text{W}}$ is the Wilson (gauge and quark) Lagrangian, 
Eqs.~(\ref{eq:WilsonAction})--(\ref{eq:WilsonQuarkAction}).
The Wilson action contains a bare gauge coupling~$g_0^2$ and a 
(dimensionless) bare mass~$m_0a$. 
In Eq.~(\ref{eq:Llat}), the interactions $O_{\text{lat}}$ are
gauge-invariant composite operators of lattice gauge and lattice
fermion fields.
The couplings~$c_O$ can be chosen in the same spirit as 
$c_{\text{SW}}$ in Eq.~(\ref{eq:SWVTX}), and we shall explain how the 
effective field theory provides a systematic framework for making a 
choice.

The right-hand side of Eq.~(\ref{eq:lat=Sym}) is a local effective 
Lagrangian (LE${\cal L}$) used for analyzing the computer output.
Its ultraviolet behavior is regulated and renormalized completely 
separately from the lattice of the left-hand side.
It is convenient to think of ${\cal L}_{\text{Sym}}$ as a continuum 
field theory.
The LE${\cal L}$ is the Lagrangian of the corresponding continuum
field theory, plus extra terms to describe discretization effects.
For lattice QCD
\begin{equation}
	{\cal L}_{\text{Sym}} = {\cal L}_{\text{QCD}} + {\cal L}_I,
	\label{eq:LEL}
\end{equation}
where ${\cal L}_{\text{QCD}}$ is the renormalized, continuum QCD 
Lagrangian,
\begin{equation}
	{\cal L}_{\text{QCD}} = \frac{1}{2g^2}\tr[F^{\mu\nu}F_{\mu\nu}]
		- \bar{q}\left({\kern+0.1em /\kern-0.65em D} + m\right)q.
	\label{eq:QCD}
\end{equation}
The renormalized gauge coupling~$g^2$ and renormalized mass~$m$ depend 
on the bare gauge coupling~$g_0^2$, the bare mass $m_0$, the $c_O$ in 
Eq.~(\ref{eq:Llat}), and the chosen renormalization scheme:
\begin{eqnarray}
	g^2 & = &     g^2(g^2_0, m_0a; c_O; \mu a), \\
	 m  & = & m_0 Z_m(g^2_0, m_0a; c_O; \mu a),
\end{eqnarray}
where $\mu$ is the renormalization point.
The continuum limit is taken for fixed $g^2$ and~$m$.
Lattice artifacts are described by operators of dimension 
$\dim{\cal O}>4$: 
\begin{equation}
	{\cal L}_I = \sum_{\cal O}
		a^{\dim{\cal O}-4} K_{\cal O}(g^2, ma; c_O; \mu a)\,
		{\cal O}_R(\mu),
		\label{eq:artifacts}
\end{equation}
where $a^{\dim{\cal O}-4}K_{\cal O}$ is a short-distance coefficient,
written with factors of $a$ so that $K_{\cal O}$ is dimensionless.
As with the renormalized couplings, these short-distance
coefficients depend on all couplings of the lattice action.

The renormalized operators ${\cal O}_R$ are sensitive to long distances
only.
In particular, they do {\em not} depend on the short distance~$a$.
Multiplying matrix elements of ${\cal O}_R$ with their coefficients, 
one finds terms of order $(pa)^{\dim{\cal O}-4}$, where $p$ is a 
typical momentum, and $\dim{\cal O}-4>0$.
For hadrons consisting of quarks and gluons, $p\sim\Lambda$.
By assumption $\Lambda a$ is small, so one can treat the lattice 
artifacts in ${\cal L}_I$ as perturbations.
To do so, one can pass to the interaction picture driven by 
${\cal L}_{\text{QCD}}$, and in this way develops a series, with all 
matrix elements taken in the (continuum) eigenstates 
of~${\cal L}_{\text{QCD}}$.

In the interaction picture, one can simplify ${\cal L}_I$ by omitting 
redundant interactions.
Let us focus on the quark part of the Lagrangian, and consider the 
field redefinitions
\begin{equation}
	q\mapsto q + a^{\dim X}\varepsilon_X Xq, \quad
	\bar{q}\mapsto \bar{q} + a^{\dim X}\bar{\varepsilon}_X\bar{q}X,
	\label{eq:change}
\end{equation}
where $X$ is an arbitrary gauge-{\em co}variant operator, and
$\varepsilon_X$ and $\bar{\varepsilon}_X$ are free parameters.
Equation~(\ref{eq:change}) simply changes the integration variables 
of the functional integral.
On-shell matrix elements, which are the integrals themselves, do not 
change.\cite{Luscher:1984xn}
Since the interaction picture is being driven 
by~${\cal L}_{\text{QCD}}$, the mass shell in question is that of QCD, 
even though we have not yet solved for the hadron masses. 

A trivial example is if $X$ is a constant.
Then Eq.~(\ref{eq:change}) changes the normalization of the fields~$q$
and~$\bar{q}$, so one concludes that on-shell matrix elements do not 
depend on how the field is normalized.
In other cases, the field redefinition acting on~${\cal L}_{\text{QCD}}$
induces higher-dimension terms, like those in~${\cal L}_I$.
Thus,
\begin{equation}
	{\cal L}_I \mapsto {\cal L}_I + \sum_X a^{\dim X}\left[
		\bar{\varepsilon}_X\bar{q}X({\kern+0.1em /\kern-0.65em D} + m_q)q +
		\varepsilon_X\bar{q}(-\overleftarrow{\kern+0.1em /\kern-0.65em D} +
		m_q)Xq \right] .
	\label{eq:changeLI}
\end{equation}
Similarly, the redefinition acting on terms in~${\cal L}_I$ changes 
terms of even higher dimension.
In summary, Eq.~(\ref{eq:change}) amounts to changing certain
coefficients in~${\cal L}_I$.
Since the changes are arbitrary, the corresponding operators
have no effect on on-shell matrix elements.
When using the effective field theory to describe the underlying theory,
their coefficients may be set according to convenience, so they are
called redundant.
They are easy to identify, because they vanish by the equations of
motion of~QCD.

Let us illustrate with dimension-five operators.
There are no pure gauge operators, but there are two linearly 
independent quark operators, namely
\begin{eqnarray}
	{\cal O}_5  & = & i\bar{q} \sigma_{\mu\nu}F^{\mu\nu} q, \\
	{\cal O}'_5 & = & 2\bar{q} D^2 q. \label{eq:D2}
\end{eqnarray}
The second of these can be re-written as
\begin{equation}
	{\cal O}'_5 = {\cal O}_5 +
		2\bar{q}{\kern+0.1em /\kern-0.65em D}
		\left({\kern+0.1em /\kern-0.65em D} +  m \right)q - 
			2m \bar{q}{\kern+0.1em /\kern-0.65em D}q .
\end{equation}
These three terms are, in order, the other dimension-five operator, 
a redundant operator, and something proportional to the kinetic term 
in~${\cal L}_{\text{QCD}}$.
The last can be absorbed into the field normalization of~$q$ and a 
redefinition of~$m$.
Thus, ${\cal O}_5$ suffices to describe all on-shell dimension-five
effects:
\begin{equation}
	{\cal L}_I = aK_{\sigma\cdot F}\, 
		\bar{q}i\sigma_{\mu\nu}F^{\mu\nu}q + \cdots, 
	\label{eq:sigmaF}
\end{equation}
where the ellipsis denotes terms of dimension six and higher.

The vector and axial vector currents can be described along a similar 
lines.
Consider, for example, the flavor-changing transition $s\to u$.
The lattice currents take the form
\begin{eqnarray}
	V^\mu_{\text{lat}} & = & \bar{\psi}_ui\gamma^\mu\psi_s -
		ac_V{\partial_\nu}_{\text{lat}}\bar{\psi}_u\sigma^{\mu\nu}\psi_s + 
		\sum_{O_V} a^{\dim O_V-3} c_{O_V} O_V^\mu ,
	\label{eq:Vlat} \\
	A^\mu_{\text{lat}} & = & \bar{\psi}_ui\gamma^\mu\gamma_5\psi_s +
		ac_A \partial^\mu_{\text{lat}} \bar{\psi}_ui\gamma_5\psi_s + 
		\sum_{O_A} a^{\dim O_A-3} c_{O_A} O_A^\mu ,
	\label{eq:Alat}
\end{eqnarray}
which are general expressions with the right quantum numbers.
In the Symanzik effective field theory these currents are described 
by\cite{Luscher:1996sc}
\begin{eqnarray}
	V^\mu_{\text{lat}} & \doteq & \bar{Z}_V^{-1} {\cal V}^\mu -
		aK_V \partial_\nu \bar{u} \sigma^{\mu\nu}s + \cdots ,
	\label{eq:LEV} \\
	A^\mu_{\text{lat}} & \doteq & \bar{Z}_A^{-1} {\cal A}^\mu +
		aK_A \partial^\mu \bar{u}i          \gamma_5 s + \cdots ,
	\label{eq:LEA}
\end{eqnarray}
where the ellipsis denotes operators of dimension four and higher,
$\bar{Z}_J^{-1}$ and $K_J$ are short-distance coefficients, and
\begin{eqnarray}
	{\cal V}^\mu & = & \bar{u}i\gamma^\mu        s ,
	\label{eq:V} \\
	{\cal A}^\mu & = & \bar{u}i\gamma^\mu\gamma_5s ,
	\label{eq:A}
\end{eqnarray}
are the vector and axial vector currents in continuum~QCD.
Like the short-distance coefficients in the effective Lagrangian,
$\bar{Z}_J$ and $K_J$ are functions of $g^2$ and $ma$, the lattice
couplings $c_O$, and the parameters~$c_{O_J}$.
Further dimension-four operators may be omitted from 
Eqs.~(\ref{eq:LEV}) and~(\ref{eq:LEA}), because they are linear 
combinations of those listed and others that vanish by the equations 
of motion.

Like the terms of dimension five and higher in~${\cal L}_I$, 
the dimension-four currents can be treated as perturbations.
Thus, the effective field theory says that
\begin{eqnarray}
	\bracket{f_{\text{lat}}}{\bar{Z}_AA^\mu_{\text{lat}}}{i_{\text{lat}}} & = &
		\bracket{f}{{\cal A}^\mu}{i} + 
		a \bar{Z}_AK_A \partial^\mu \bracket{f}{\bar{u}i\gamma_5s}{i}
	\label{eq:SymAxial} \\
		& + & a K_{\sigma\cdot F}\int d^4x\,
		\bracket{f}{T\,{\cal O}_5 {\cal A}^\mu }{i} + O(a^2),
	\nonumber
\end{eqnarray}
and similarly for the vector current. 
The crucial feature of Eq.~(\ref{eq:SymAxial}) is that, while the 
states on the left-hand side are eigenstates of lattice gauge theory, 
those on the right-hand side are eigenstates of continuum~QCD.

The Symanzik effective field theory can be justified in to all orders 
in perturbation theory (in the gauge coupling).
In the late 1980's Reisz proved a power counting 
theorem\cite{Reisz:1987da} that enabled him to set up a version of 
BPHZ (Bogoliubov-Parasiuk-Hepp-Zimmermann) renormalization tailored to 
the Feynman diagrams of lattice perturbation theory.
With mild assumptions on the gluon and quark propagators,%
\footnote{Note that Wilson fermions satisfy the assumptions, but 
staggered fermions do not.}
he showed that loop integrals with external momenta $\{p\}$ 
and loop momenta $\{k\}$ can be written\cite{Reisz:1987px}
\begin{equation}
	\int\prod_{i=1}^l\frac{d^4k_i}{(2\pi)^4}\,{\cal I}(\{p\},\{k\}) =
		I_R + I_U(\{p\}) + O(ap,am),
	\label{eq:Reisz}
\end{equation}
where $I_R$ denotes parts that are absorbed when renormalizing the 
gauge coupling and the quark masses.
The other term~$I_U(\{p\})$ does not depend on~$a$ or on the 
coefficients~$c_O$ of higher-dimension lattice interactions.
Thus, after renormalization, lattice perturbation theory is
universal:\cite{Reisz:1988kk} the continuum limit does not depend on 
the discretization.
The remainder terms can be developed further with BPHZ 
oversubtractions of the loop integrands.
In this way one can develop any amplitude's renormalized perturbation 
series, including contributions suppressed by powers of $a$, to any 
order desired.
Similarly, one can develop the renormalized perturbation series
generated by Symanzik's LE${\cal L}$.
To lend rigor to the Symanzik theory, one could imagine defining the 
LE${\cal L}$ on an infinitesimally fine lattice, and using Reisz's 
methods to extract the universal part, and also to define the operator 
insertions of~${\cal L}_I$.
But in practice any method of regulating and renormalizing the 
ultraviolet will do.

Although it is only fully justified in perturbation theory, the 
Symanzik effective field theory is believed to hold at a 
non-perturbative level as well.
At short distances, small instantons make contributions that should 
be lumped into the short-distance coefficients.
But they come in with a high power of $a$ and are presumably negligible.
Long-distance phenomena, such as confinement, are more mysterious.
But this mystery is bundled into the QCD term in the LE${\cal L}$, 
not ${\cal L}_I$.
Indeed, if the Symanzik formalism were to fall apart, it would be 
hard to understand why other short-distance methods of~QCD work so 
well to explain measurements of deeply inelastic scattering, jet 
cross sections, or $B$~decays.

By calculating on-shell matrix elements on both sides of, say, 
Eq.~(\ref{eq:SymAxial}) in renormalized perturbation theory, 
one can read off the short-distance coefficients in~${\cal L}_I$.
In perturbation theory, let
\begin{equation}
	K_{\cal O}(g^2, ma; c_O; \mu a) = g^{2n}
		\sum_{l=0}^{\infty} g^{2l} K_{\cal O}^{[l]}(ma; c_O; \mu a),
\end{equation}
where $n\ge0$: in some cases the coefficients vanish at the tree 
level.
From the justification of the LE${\cal L}$ given above, it is 
natural to use the renormalized coupling in this series.
An especially transparent choice of $g^2$ is something physical, at 
least in perturbation theory, for example the coupling that appears 
in scattering of two quarks of different flavor.
Then from familiar properties of perturbation theory
\begin{equation}
	K_{\cal O}^{[l]}(ma; c_O; \mu a) = \sum_{r=0}^l 
		K_{\cal O}^{[l,r]}(ma; c_O) \;(\ln \mu a)^r.
\end{equation}
The $\mu$ dependence must arise in the right way to cancel 
the $\mu$~dependence of the operators~${\cal O}_R$ in~${\cal L}_I$.

In the discussion so far, the Symanzik effective field theory is 
simply a theory of cutoff effects.\cite{Symanzik:1979ph}
It applies no matter how the lattice couplings $c_O$ are chosen.
The default is $c_O=0$: most higher-dimension lattice 
interactions are omitted from the Lagrangian inside the computer.
At this level, the Symanzik theory yields, with great sophistication, 
the result that on-shell matrix elements suffer lattice artifacts 
suppressed by powers of~$a$.

A lattice gauge theory with smaller short-distance coefficients in the
LE${\cal L}$ would yield better estimates of continuum~QCD.
This is where the Symanzik formalism becomes especially powerful, 
because one can demand $K_{\cal O}=0$, and solve for the lattice 
couplings~$c_O$.\cite{Symanzik:1983dc}
This is called the Symanzik improvement program.
A~key feature is that if a coefficient~$K_{\cal O}$ (or its expansion 
coefficient~$K_{\cal O}^{[l]}$) is made to vanish for one observable, 
then the effective field theory set up shows that it vanishes for all 
observables.

Of course, it is impractical to solve the equations $K_{\cal O}=0$ in 
any kind of generality.
At the very least, one must truncate in (scaling) dimension.
For QCD, asymptotic freedom guarantees that the anomalous dimensions 
are small, so the scaling dimension is not much different from the 
classical dimension.
At dimension five there is only one term in~${\cal L}_I$, 
cf.\ Eq.~(\ref{eq:sigmaF}), and with the Wilson action, 
$K_{\sigma\cdot F}\neq0$.
But, following Sheikholeslami and Wohlert,\cite{Sheikholeslami:1985ij}
one can write down a discretization of~${\cal O}_5$,
\begin{equation}
	O_5 = i\bar{\psi}\sigma_{\mu\nu}G^{\mu\nu}\psi,
\end{equation}
where $G^{\mu\nu}$ is a combination of SU(3) $U$ matrices,
defined in Eq.~(\ref{eq:Umx}), such that the naive $a\to0$ limit
gives~$F^{\mu\nu}$.
The Sheikholeslami-Wohlert Lagrangian
\begin{equation}
	{\cal L}_{\rm SW} = {\cal L}_{\text{W}} + 
		\case{1}{4} c_{\text{SW}} O_5.
	\label{eq:SWaction}
\end{equation}
The Symanzik coefficient $K_{\sigma\cdot F}$ is, of course, very 
sensitive to the coupling~$c_{\text{SW}}$.
At the tree level
\begin{equation}
	K_{\sigma\cdot F}^{[0]}(ma) = \case{1}{4}(1-c_{\text{SW}}).
	\label{eq:Ctree}
\end{equation}
Thus, as in Eq.~(\ref{eq:SWVTX}), setting $c_{\text{SW}}=1$ in the 
computer calculations reduces the leading lattice artifact.

In the foregoing discussion, the short-distance coefficients are 
written as functions of~$ma$, to emphasize that the primary goal of 
the effective field theory is to separate long- and short-distance 
scales.
For light quarks, however, it is safe to expand the short-distance 
coefficients in powers of~$ma$.
One can then use this expansion and the ordering of~${\cal L}_I$ by 
dimension to develop asymptotic expansions in powers of~$a$.
This is the most widely recognized application of the Symanzik 
effective field theory.
Indeed, many reviews of the formalism skip over the scale-separation 
aspect and focus entirely on the lattice-spacing expansion.

To eliminate all terms of order~$a$, the improvement program is 
relatively straightforward.
One needs only the first two terms in the small~$a$ expansion of the 
normalization factors
\begin{equation}
	\bar{Z}_J(ma) = Z_J\left[1 + b_J \case{1}{2}(m_u+m_s)a \right] 
		+ O(m^2a^2).
	\label{eq:ZJma}
\end{equation}
For heavy quarks, the expansion in $m_Qa$ only makes sense if $m_Qa\ll1$.
As discussed in Sec.~\ref{sec:why}, this situation is not easy to 
attain for the $b$ quark.
We shall return to further aspects of the Symanzik LE${\cal L}$ for 
heavy quarks in Sec.~\ref{sec:hqet}.
But for light quarks Eq.~(\ref{eq:ZJma}) is sensible and useful.
In the same vein, it is consistent to replace $K_{\sigma\cdot F}$ 
and~$K_J$ with their values at $m_qa=0$.
This set of choices is called $O(a)$ improvement.

At the tree level, for currents given by just the first two terms of 
Eqs.~(\ref{eq:Vlat}) and~(\ref{eq:Alat}), respectively,
one finds Eq.~(\ref{eq:Ctree})
\begin{eqnarray}
	Z_J^{[0]} & = & 1,         \label{eq:Ztree} \\
	b_J^{[0]} & = & 1,         \label{eq:btree} \\
	K_J^{[0]} & = & c_J^{[0]}. \label{eq:Ktree}
\end{eqnarray}
The one-loop corrections to the $O(a)$ terms have been calculated, 
with the improved lattice Lagrangian~${\cal L}_{\text{SW}}$,
for the Lagrangian itself\cite{Wohlert:1985up} 
and for the currents.\cite{Luscher:1996lw,Sint:1997fp}
The results are not especially informative, except\cite{Sint:1997fp}
\begin{eqnarray}
	K_V^{[1]} & = & c_V^{[1]} + C_F 0.01225, \label{eq:KVloop} \\
	K_A^{[1]} & = & c_A^{[1]} + C_F 0.00568. \label{eq:KAloop}
\end{eqnarray}
Thus, one can obtain the desired improvement condition $K_V=K_A=0$
by setting $c_V=0-g^2C_F 0.01225+O(g^4)$, $c_A=0-g^2C_F 0.00568+O(g^4)$.

An important development of recent years are methods for computing the 
short-distance coefficients, through $O(a)$, non-perturbatively.
There are two key ingredients.
The first is chiral symmetry.
In 1985 it was observed\cite{Bochicchio:1985za} that, as $ma\to0$, 
both $Z_V$ and $Z_A$ can be computed non-perturbatively by imposing the
chiral Ward identities.
For flavor conserving currents, $\bar{Z}_V$ is simply the factor that 
normalizes the flavor charge.
Then, because chiral Ward identities relate vector and axial vector 
correlation functions, they fix~$Z_A$.
The second key ingredient is the Symanzik effective field theory, 
which provides a framework for imposing chiral symmetry through $O(a)$.%
\cite{Luscher:1996sc,Jansen:1996ck}

To compute the $O(a)$ terms, one can exploit the partially conserved 
axial current (PCAC).\cite{Jansen:1996ck}
In continuum QCD the PCAC relation is
\begin{equation}
	\partial_\mu {\cal A}^\mu - (m_s+m_u){\cal P} = 0,
	\label{eq:PCAC}	
\end{equation}
where the pseudoscalar density ${\cal P}=i\bar{u}\gamma_5s$.
The implication of writing Eq.~(\ref{eq:PCAC}) as an operator equation 
is that it holds for all combinations of initial and final states.
A short manipulation with the Symanzik effective theory shows that
\begin{eqnarray}
	\langle f|\bar{Z}_A{\partial_\mu}_{\text{lat}}A^\mu_{\text{lat}} 
		& - & (m_s+m_u)\bar{Z}_PP_{\text{lat}}|i\rangle =
	\label{eq:PCAClat} \\ & & \hspace{-5em}
		a\bar{Z}_AK_A \partial^2\bracket{f}{{\cal P}}{i}
		+ 2aK_{\sigma\cdot F} 
		\bracket{f}{\bar{u}\sigma\cdot F\gamma_5s}{i} + O(a^2),
	\nonumber
\end{eqnarray}
for any initial and final states.
By choosing three combinations of states, one may use one to eliminate 
$(m_u+m_s)\bar{Z}_P/\bar{Z}_A$, and then use the other two to adjust 
$c_A$ (introduced in Eq.~(\ref{eq:Alat})) and $c_{\text{SW}}$ so that 
the right-hand side vanishes, as desired for continuum QCD.
Because the renormalized mass drops out, the resulting conditions on 
$c_A$ and $c_{\text{SW}}$ do not depend on the renormalization scheme 
of the effective theory.
In a non-perturbative calculations, however, the $O(a^2)$ terms are
ever present, so $c_A$ and $c_{\text{SW}}$ are determined only with
an accuracy of order~$\Lambda a$, coming from the $O(a^2)$ matrix
elements on the right-hand side of Eq.~(\ref{eq:PCAClat}).
The improvement coefficient of the vector current, $c_V$, and the 
normalization constants $Z_J$ and $b_J$ are then determined by 
imposing Ward identities.

The non-perturbative calculation of $c_{\text{SW}}$ is 
straightforward,\cite{Luscher:1996ax} and the value obtained 
is relatively insensitive to details such as the states used 
in Eq.~(\ref{eq:PCAClat}).
It also agrees well with perturbation theory.\cite{Wohlert:1985up}
On the other hand, the non-perturbative calculation of~$c_A$ is not
so straightforward---two different groups find marginal agreement,%
\cite{Luscher:1996ax,Bhattacharya:2001pn} and it has been
found\cite{Collins:2001mm} to depend on the difference
operator~$\partial^\mu_{\text{lat}}$.
These difficulties are, perhaps, to be expected, since in
Eq.~(\ref{eq:PCAClat}) $K_A$ is multiplied with the small
quantity~$\partial^2\bracket{f}{{\cal P}}{i}$.
The non-perturbative results also do not agree well with perturbation 
theory.
The situation for $c_V$ is similarly unsettled.\cite{Bhattacharya:2001pn}
All estimates for $c_A$ and $c_V$ yield small values,
but they multiply large matrix elements.
For example,\cite{Collins:2001mm} in computing $f_\pi$ the small
correction is $c_Aam_K^2/m_s$, and $m_K^2/m_s=2.5$~GeV.

In the non-perturbative improvement program, symmetries of continuum QCD
were imposed to determine improvement coefficients of the lattice action
and currents.
Another example of the interplay of symmetry and Symanzik's theory comes
in the lattice calculation of the kaon bag parameter~$B_K$, which arise
in the theory of $K^0$-$\bar{K}^0$ mixing.
$B_K$~is defined by
\begin{equation}
	\langle\bar{K}|{\cal Q}|K\rangle = \case{8}{3}m_K^2f_K^2B_K
\end{equation}
where the $\Delta S=2$ four-quark operator
\begin{equation}
	{\cal Q} = \bar{s}\gamma^\mu(1-\gamma_5)d\,
		 \bar{s}\gamma_\mu(1-\gamma_5)d,
\end{equation}
and the kaon decay constant $f_K\approx160$~MeV is defined through
\begin{equation}
	\langle 0|{\cal A}^\mu|K\rangle = ip_K^\mu f_K.
\end{equation}
To calculate $f_K$ and $B_K$ one starts with $A^\mu_{\text{lat}}$ and 
a lattice approximant to~${\cal Q}$, which we will call~$Q^{(6)}$.
In the Symanzik effective field theory, the lattice axial vector
current is described by Eq.~(\ref{eq:LEA}) and $Q^{(6)}$ by
\begin{equation}
	Q^{(6)}_i \doteq Z_{ij}^{-1}\left[{\cal Q}_j^{(6)} 
		+ a C_{jk}{\cal Q}_k^{(7)}\right],
	\label{eq:Q6}
\end{equation}
where $Z_{ij}^{-1}$ and $ C_{jk}$ are short-distance coefficients.
The chiral flavor symmetry group in lattice gauge theory is usually 
smaller than ${\rm SU}(n_f)\times{\rm SU}(n_f)$, so one must allow for 
mixing between the target operator~${\cal Q}$ and several other 
operators, indexed by the subscripts~$i$, $j$,~$k$.
Similarly, the leading lattice spacing effects are described by 
several operators~${\cal Q}_k^{(7)}$.
As usual, the operators on the right-hand side of Eq.~(\ref{eq:Q6})
are defined in continuum~QCD.
Thus, in analogy with Eq.~(\ref{eq:SymAxial}), the effective theory 
says
\begin{eqnarray}
	\langle\bar{K}_{\text{lat}}|Z_{ij}Q_j|K_{\text{lat}}\rangle & = &
		\langle\bar{K}|{\cal Q}^{(6)}_i|K\rangle + 
		a C_{ik} \langle\bar{K}|{\cal Q}^{(7)}_j|K\rangle
	\label{eq:SymBK} \\
		& + & a K_{\sigma\cdot F}\int d^4x\,
		\langle\bar{K}|T\,{\cal Q}^{(6)}_i\sum_{q=d,\,s}{\cal O}_5(x) 
		|K\rangle + O(a^2),
	\nonumber
\end{eqnarray}
where the states on the left-hand side are eigenstates of the lattice
theory, whereas the states on the right-hand side are continuum QCD
eigenstates.
For other four-quark operators the effective theory description also 
contains operators of dimension less than six, multiplied by $1/a$.
Such ``power-law divergences'' are absent here, because all operators 
must have $\Delta S=2$.

Equation~(\ref{eq:SymBK}) holds for all formulations of lattice
fermions, but it becomes interesting for Kogut-Susskind (or staggered)
fermions.
One Kogut-Susskind field produces four flavors in the continuum limit,
so the lists of operators ${\cal Q}_i^{(s)}$ contain various flavor
combinations.
The drawback of four flavors is ameliorated by a remnant of exact chiral 
symmetry as $m\to0$, cf.\ Sec.~\ref{sec:chiral}.
The symmetry requires $K_{\sigma\cdot F}$ to vanish as~$m_sa$.
On the other hand, the lattice artifacts of the operator do not vanish:
the short-distance coefficients~$C_{ik}$ are non-zero.
The matrix elements~$\bracket{\bar{K}}{{\cal Q}^{(7)}_j}{K}$
appearing in Eq.~(\ref{eq:SymBK}) {\em do\/} vanish, however, 
by the flavor symmetry of (four-flavor) continuum QCD.%
\cite{Sharpe:1993ng,Sharpe:1994dc,Luo:1996vt}
Thus, the leading lattice artifacts in the lattice calculation of
$\case{8}{3}m_K^2f_K^2B_K$ is of order~$a^2$.
A~similar argument applies to the matrix elements of the axial vector
current giving~$f_K$, so the leading lattice artifact in~$B_K$, when 
calculated with staggered quarks, is also of order~$a^2$.

In the last few years, there has been spectacular development in the 
understanding of chiral lattice gauge theories.%
\cite{Niedermayer:1998bi,Neuberger:1999ry,Luscher:1999mt}
For vector-like gauge theories like QCD, these developments have led 
to methods with either very small violations of chiral 
symmetry\cite{Shamir:1993zy} or essentially no violations of chiral 
symmetry.\cite{Wiese:1993cb,Hasenfratz:1998ri,Neuberger:1997fp}
When these methods are used, the leading lattice spacing errors are of 
order~$a^2$.
Indeed, one of the methods---domain-wall 
fermions\cite{Shamir:1993zy,Kaplan:1992bt,Blum:1996jf}---has been
applied to~$B_K$.\cite{AliKhan:2001wr,Blum:2001xb}
These calculations also seem to have much smaller $O(a^2)$ effects
than state-of-the-art calculations\cite{Kilcup:1997ye,Aoki:1997nr}
with staggered fermions.

With any of several methods---$O(a)$ improved Wilson
fermions, staggered fermions, domain-wall fermions, overlap
fermions\cite{Neuberger:1997fp}---the leading cutoff effect is of
order~$a^2$.
To reduce them further, one is confronted with many, many operators 
of dimension six in the Lagrangian, of dimension five in the currents, 
etc.
It may not be feasible to eliminate all these effects with 
non-perturbative methods---especially in light of the difficulties 
with~$c_A$.
On the other hand, it is possible to compute these coefficients 
in perturbation theory.

The technical details of lattice perturbation theory are more 
cumbersome that in continuum QCD, as seen in the Feynman rule in 
Eq.~(\ref{eq:SWVTX}).
But, since the ultraviolet cutoff is built in, the problem is well 
suited to computer algebra.
One can generate vertices and propagators and combine them into 
diagrams automatically.\cite{Luscher:1985wf,Nobes:2001tf}
Although the lattice renders the integrals ultraviolet finite,
difficulties can arise in the infrared.
But these effects are universal, essentially by Reisz's theorem.
Thus, once the infrared has been understood for a simple lattice 
Lagrangian, the improvement interactions merely add algebraic 
complexity that a computer program can handle.
These automated methods were first introduced for pure gauge theory 
at the one-loop level,\cite{Luscher:1985wf} but are now being extended 
to QCD at the two-loop level.\cite{Nobes:2001tf}

A more controversial issue is the viability of perturbation theory 
in this context.
Because lattice gauge theory is a theoretically consistent whole,
it seems at first glance natural to use the bare coupling~$g_0^2$ 
as the expansion parameter of perturbative series.
This choice ignores the fact that one wants to connect lattice gauge 
theory (at moderate lattice spacing) to continuum~QCD.
The set up of the Symanzik effective field theory, especially as 
underpinned by Reisz's work, encourages the use of a renormalized 
coupling.
Furthermore, from a practical point of view, the bare coupling in the 
Wilson gauge action is a disastrous choice: if one compares 
perturbative and Monte Carlo calculations of short-distance quantities 
such as small Wilson loops, the agreement is very poor.
On the other hand, renormalized perturbation theory usually describes 
short-distance properties well, especially when the renormalized 
coupling is chosen according to the Brodsky-Lepage-Mackenzie (BLM) 
prescription.\cite{Brodsky:1983gc,Lepage:1993xa}
For example, the differences between one-loop BLM perturbation 
theory for the renormalization and improvement of the vector and 
axial vector currents is easily attributed to (as yet uncomputed) 
two-loop effects.\cite{Harada:2002pt}

\begin{figure}[tbp]
	\centering
 	\includegraphics[width=0.8\textwidth]{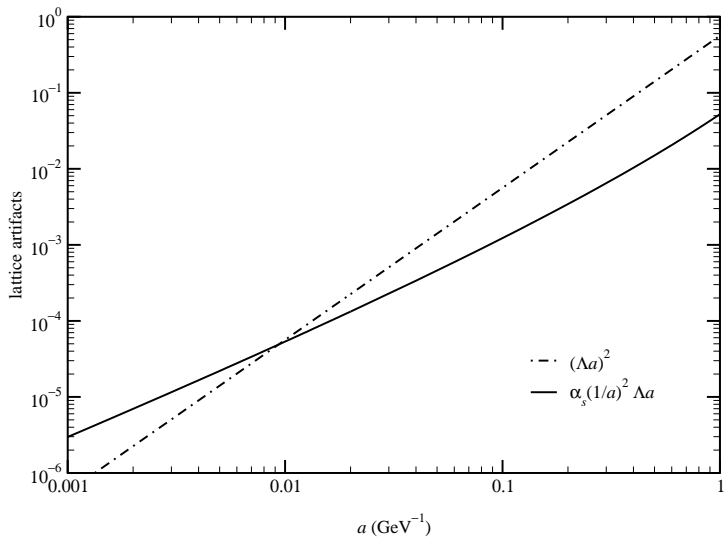}
	\caption{Comparison of the ``dominant'' lattice artifact
	$\alpha_s(1/a)^2\Lambda a$ with the ``sub-dominant''~$(\Lambda a)^2$.
	Except at unrealistically small $a$, the latter is larger.}
	\label{fig:latart}
\end{figure}

Even assuming perturbation theory is accurate, one is faced with a 
formal issue.
To illustrate it, let us assume that the we have $O(a)$ improvement 
of Wilson fermions at the one-loop level.
Then the leading lattice spacing effects (for light hadron masses) are 
of order $\alpha_s(1/a)^2\Lambda a$ and $(\Lambda a)^2$.
Formally, the former dominates as $a\to0$.
Figure~\ref{fig:latart} shows $\alpha_s(1/a)^2\Lambda a$ and 
$(\Lambda a)^2$ as a function of~$a$ for $\Lambda=750$~MeV and 
$\alpha_s(2~{\rm GeV})=0.2$.
In the range where lattice calculations can be done, 
$a>0.25~{\rm GeV}^{-1}$, the ``sub-dominant'' effect 
is an order of magnitude larger than the ``dominant'' one.
Ideally, one would have enough data to fit to both contributors.
Otherwise, one is faced with the choice of introducing a bias, 
by ignoring~$\alpha_s(1/a)^2a$ and fitting to~$a^2$, 
or introducing an error,
by ignoring~$a^2$ and fitting to~$\alpha_s(1/a)^2a$ (or~$a$). 
Figure~\ref{fig:latart} suggests that the uncertainty stemming from 
the second Ansatz is larger that the uncertainty stemming from the 
first.

In the last few years, Symanzik improvement has been a major focus of
research in lattice gauge theory.
In light of this renaissance of Symanzik's work, it is surprising that
many papers still report calculations at only one lattice spacing.
The improvement program is based on a theory of cutoff effects, which
clearly demonstrates the utility of repeating the calculation at several
lattice spacings.
To see the benefits, let us consider a simple example.
Suppose one has 100 (arbitrary) units of CPU available.
Instead of spending all 100 units on the finest possible lattice
spacing~$a_0$, one could consider coarser lattices of spacing
$a_1=a_0/\sqrt[4]{2}$ and $a_2=a_0/\sqrt{2}$.
Spending 65, 25, and 10 units of CPU at $a_0$, $a_1$, and $a_2$ would,
according to Eq.~(\ref{eq:gaugeslow}), yield comparable statistical
errors.
Compared to putting all 100 units at $a_0$, the statistical error would
be a bit larger, but only by a factor of 1.25 ($=1/\sqrt{0.65}$).
But, if all 100 units are spent at $a_0$, then one has an uncertainty,
of order $(\Lambda a)^2$ say, which requires a guess for the appropriate
$\Lambda$: 250~MeV or 2.5~GeV?
A calculation based on three (or more) lattice spacings does not require
a guess, because the added information is tantamount to a calculation of
the discretization effect.
The slightly larger statistical error seems a small price to pay.

\section{Heavy Quark Effects: Heavy-Quark Effective Theory}
\label{sec:hqet}

Some of the most interesting applications of numerical lattice gauge 
theory arise in heavy quark physics.
The main aim of experimental $B$ physics is to study flavor and $CP$
violation precisely enough to test the CKM mechanism.
To connect the CKM matrix or, indeed, other short-distance mechanisms of
flavor and $CP$ violation, one is faced with theoretical formulae
of the form
\begin{equation}
	\left( \begin{array}{c} \rm measured       \\
		\rm quantity \end{array}\right) =
	\left( \begin{array}{c} \rm kinematic      \\
		\rm factors  \end{array}\right)
	\left( \begin{array}{c} \textrm{short-distance} \\
		\rm factor   \end{array}\right)
	\left( \begin{array}{c} \rm QCD            \\
		\rm factor   \end{array}\right),
\end{equation}
where the measured quantity is a (differential) rate, and the 
kinematic factors consist of measurable momenta and (hadron) masses.
Here the short-distance factor includes wavelengths less than 
$(100~\text{GeV})^{-1}$; in the Standard Model, it consists of 
well-determined parameters (like the Fermi constant~$G_F$) and the 
less well-determined CKM matrix.
The QCD factor, as a rule, boils down to a hadronic matrix element.
In the case of $|V_{ud}|$ and $|V_{us}|$, isospin and SU(3) symmetries
bring the QCD under control, but in other cases lattice calculations
are essential.\cite{Beneke:2002ks,Hashimoto:2000bk}
In some $B$ decays heavy-quark symmetry provides some control, but
hadronic matrix elements still appear in contributions at the $1/m_Q$ 
level, which, generically, are 10\%~effects.

Measured in lattice units, the bottom and charmed quarks' masses can 
easily be large.
Even if $\Lambda a\sim0.1$--0.3, so that the Symanzik effective field 
theory works well for gluons and light quarks, then $m_ba\sim1$--2 and 
$m_ca$ about a third of that.
For this reason it is frequently (but incorrectly) stated that heavy
quarks cannot be directly accommodated by a lattice.
This observation overlooks the physical fact that the heavy-quark mass
scale is far removed from the QCD scale and that, consequently, the
dynamics of heavy-quark systems simplify.
Nevertheless, it is fair to say that lattice spacing effects are more
challenging for heavy quarks than for light quarks.
So, in this section, we discuss how to classify, control, and minimize
discretization uncertainties of heavy quarks.

The key is to make use of effective field theories for heavy-quark
systems.
Indeed, from the inception of non-relativistic QCD (NRQCD)%
\cite{Caswell:1986ui,Lepage:1987gg,Lepage:1992tx}
and heavy-quark effective theory (HQET),%
\cite{Eichten:1987xu,Eichten:1990zv,Eichten:1990vp}
these effective field theories have been used to treat heavy quarks
in lattice gauge theory.
Indeed, the early papers\cite{Caswell:1986ui,Lepage:1987gg,%
Eichten:1987xu,Eichten:1990zv} inspired the development of HQET and 
NRQCD with continuum ultraviolet regulators, 
as methods for understanding heavy-light
hadrons\cite{Grinstein:1990mj,Georgi:1990um,Luke:1990eg}
and quarkonia.\cite{Bodwin:1992bl}
More recently it has been shown how to use the continuum effective 
field theories to understand the heavy-quark discretization effects 
of Wilson fermions.%
\cite{El-Khadra:1997mp,Kronfeld:2000ck,Harada:2001hl,Harada:2001hh}

To make a connection with Sec.~\ref{sec:sym}, let us start with Wilson 
QCD and examine how the Symanzik theory breaks down when $ma\not\ll1$.
A~key to the Symanzik LE${\cal L}$ is that the leading dynamics are
those of ${\cal L}_{\text{QCD}}$, while the corrections~${\cal L}_I$
are small.
When $ma\not\ll1$, this split into large+small no longer holds.
First, the expansion of short-distance coefficients in small~$ma$ is 
no longer admissible.
Furthermore, ${\cal L}_I$ contains terms that scale as $ma$ to 
some power.
Consider, in particular,
\begin{equation}
	{\cal L}_I = \cdots + \sum_X a^{\dim X-1}\,\sum_{n=3}^\infty
		K_X^{(n)}\,\bar{q}X\sum_{\mu=1}^4(-\gamma_\mu D_\mu a)^nq
		+ \cdots,
	\label{eq:badO}
\end{equation}
which describe deviations from Lorentz (or Euclidean) invariance.
(They do respect hypercubic rotations.)
At each $n$, the term with $\mu=4$ is not small, 
because $(-\gamma_4D_4a)^n\sim(ma)^n$.
But one can cull $\gamma_4D_4$ from~${\cal L}_I$ by applying the 
equation of motion,\cite{Aoki:2001ra}
\begin{equation}
	-\gamma_4D_4q=(\bbox{\gamma}\cdot\bbox{D}+m)q.
\end{equation}
Repeated application of the equation of motion to yields
$a^n(\bbox{\gamma}\cdot\bbox{D}+m)^n$ and (nested) commutators of $D_4$
and~$\bbox{D}$.
The commutators do not lead to the heavy-quark mass, but to the gluon
field strength and derivatives thereof.
Thus, all large terms come from expanding
$a^n(\bbox{\gamma}\cdot\bbox{D}+m)^n$ and collecting
$(ma)^{n-r}\bar{q}X(\bbox{\gamma}\cdot\bbox{D})^rq$ into
a new coefficient for $\bar{q}X(\bbox{\gamma}\cdot\bbox{D})^rq$.
If $X=1$ and $r=0$~or~1, the coefficients in~${\cal L}_{\text{QCD}}$ 
are modified, and the LE${\cal L}$ takes the form
\begin{equation}
	{\cal L}_{\rm Sym} =
	{\cal L}_{\rm gauge} +
	\bar{q}\left(\gamma_4D_4 +
		\sqrt{\frac{m_1}{m_2}}\bbox{\gamma}\cdot\bbox{D} + m_1\right)q + 
	{\cal L}'_I,
	\label{eq:badSym}
\end{equation}
where the coefficients $m_1$ and $\sqrt{m_1/m_2}$ result
from coalescing all terms multiplying $\bar{q}q$ and
$\bar{q}\bbox{\gamma}\cdot\bbox{D}q$, respectively.
The notation is taken from the energy of a quark with small 
momentum~$\bbox{p}$:
\begin{equation}
	E(\bbox{p}) = m_1 + \frac{\bbox{p}^2}{2m_2} + O(\bbox{p}^4).
\end{equation}
Below $m_1$ is called the rest mass, and $m_2$ the kinetic mass.
With these rearrangements, the operators in ${\cal L}'_I$ now all 
yield powers of~$\Lambda a$, not~$ma$, so they still can be treated as 
operator insertions.

Equation~(\ref{eq:badSym}) rests on the same foundation as
Eq.~(\ref{eq:LEL}); only the split between large and small is different,
reflecting the new situation $ma\not\ll1$.
For terms in Eq.~(\ref{eq:badO}) with $X\neq1$, the rearrangement can be
absorbed into the short-distance coefficients of ${\cal L}'_I$.
For the Wilson fermion action, none of these coefficients (except $m_1$)
is unbounded as $ma\to\infty$.%
\cite{Eichten:1987xu,El-Khadra:1997mp,Kronfeld:2000ck}
It is important, when developing an improvement program for heavy 
quarks, not to sacrifice this property.
In fact, the currents used in the $O(a)$ improvement program 
discussed in Sec.~\ref{sec:sym} do not work well for heavy quarks.
The foregoing analysis even suggests a suitable improvement program: 
the large-mass behavior of ${\cal L}'_I$ remains well-behaved if one 
mimics it and, hence, omits from Eq.~(\ref{eq:Llat}) operators with 
extra time difference operators.\cite{El-Khadra:1997mp}
Similarly, the corrections to the currents should not have any time 
derivatives at all.\cite{Harada:2001hl,Harada:2001hh}

Unfortunately, unless $m_1=m_2$ the LE${\cal L}$ is no longer
``QCD plus small corrections,'' because the normalization of the
spatial kinetic energy is incorrect by the factor $\sqrt{m_1/m_2}$.
In free field theory, for Wilson fermions,
\begin{equation}
	\frac{m_1}{m_2} = 1 - \case{2}{3}m_1^2a^2 +
		\case{1}{2}m_1^3a^3 + \cdots,
	\label{eq:m1m2}
\end{equation}
with no term of order~$ma$.
This feature persists to all orders in perturbation theory.%
\cite{Mertens:1998wx,Harada:2001hl}
The deviations from the desired $m_1/m_2=1$ can be sizable.
For $ma=0.8$ the two terms shown are $-0.46$ and $+0.26$.
Such strong deviations from continuum QCD remain when the gauge
interaction is turned on.
The small $ma$ expansions of other short-distance quantities, for 
example the normalization factors of the currents, also break down as 
soon as $ma\not\ll1$.

There are three remedies to this problem.
First, one can take $ma\ll1$, so that the rearrangement discussed 
above is unnecessary.
Second, one can introduce another parameter to the lattice fermion 
action, so that $\bar{q}\gamma_4D_4q$ and 
$\bar{q}\bbox{\gamma}\cdot\bbox{D}q$ can be normalized separately.
Third, one can note that it is not lattice gauge theory that breaks 
down when $ma\not\ll1$ but the Symanzik effective field theory.
This leaves open the possibility that other tools can be used to 
control the discretization effects of heavy quarks.

Let us start by considering $ma\ll1$.
In principle, this is fine, because one can expand the short-distance 
coefficients in~$ma$, and Eq.~(\ref{eq:badSym}) reverts to 
``QCD plus small corrections'' as in Sec.~\ref{sec:sym}.
In practice there are serious difficulties.
As discussed in Sec.~\ref{sec:mc} it will not be possible to reduce 
$a$ enough to make $m_ba\ll1$ for many, many years.
Another way to reduce $ma$ is to reduce the heavy quark mass.
But if $m<m_b$, one must use the heavy-quark expansion to extrapolate
back up to~$m_b$.\cite{Gavela:1988cf,Bernard:1988dy}
The simultaneous requirements $ma\ll1$ and $\Lambda/m\ll1$ fight 
against each other, making it hard, on accessible lattices, to control 
both systematic errors, not to mention the crosstalk between them.
To obtain $ma<0.5$, one must take $m<m_c$, 
sometimes as small as 500~MeV.
It is not clear whether this regime can be connected back to $m_b$
through the $1/m$ expansion.
To reach larger quark masses, $m\sim1.5$~GeV, 
$ma$ is sometimes as large as 0.7--0.8.
Then discretization effects of order $(ma)^2$ and $(ma)^3$ are large.
Finally, the $(ma)^n$ errors are amplified in the $1/m$ extrapolation, 
but no one has a solid idea for estimating how much.
Concerns of this kind have been voiced before; 
Sommer\cite{Sommer:1994fr} and Wittig\cite{Wittig:1997tr} have 
insisted on taking $ma\to0$ before carrying out any extrapolation 
in~$1/m$.
Then lattice-spacing and heavy-quark effects are decoupled, and the 
main drawback of their analyses (of data in the literature), 
is that the heavy quark mass is too small.

A variation on this theme is to set up a lattice gauge theory with
different temporal and spatial lattice spacings, $a_t$ and~$a_s$.
Such lattices are called anisotropic.
The hope\cite{Klassen:1999al} is that the heavy-quark mass appears in
short-distance coefficients as~$ma_t$, but not as~$ma_s$.
Then one could take $a_t/a_s\sim\Lambda/m_b$, expand the 
short-distance coefficients in $ma_t$, and determine the improvement 
coefficients non-perturbatively.\cite{Klassen:1999al}
Unfortunately, there is no proof that $ma_s$ does not arise, and, for 
Klassen's choice of the lattice couplings, it does.\cite{Harada:2001ei}
When $ma_s$ does appear in coefficients, one cannot take 
$a_t/a_s\sim\Lambda/m_b$, and another remedy is needed.%
\footnote{Even if they do not tame heavy-quark cutoff effects, 
anisotropic lattices still can be useful for reducing statistical 
uncertainties, by providing more timeslices in the region of Euclidean 
time where one state saturates Eq.~(\ref{eq:C2exp}).%
\cite{Collins:2001pe}}

The second remedy is to modify the lattice gauge theory so that the
temporal and kinetic terms are separately adjustable.
A simple way to this was introduced by El-Khadra
et~al.\cite{El-Khadra:1997mp}
Then one could adjust the underlying lattice parameters so that 
in the LE${\cal L}$, $m_1\equiv m_2$.
The adjustment can be made non-perturbatively, by forcing the rest 
mass and the kinetic mass of a hadron to be the same.
It works well,\cite{Sroczynski:1999he} but has not been widely used
in heavy-quark phenomenology.
In this approach the $ma$ dependence of the short-distance 
coefficients is not as simple as for light quarks.
But at least it is possible to set $m=m_b$, circumventing the
heavy-quark extrapolation.

The third remedy is to set up the calculation so that heavy-quark 
methods (HQET or NRQCD) and lattice gauge theory work together.
There are two ways to go about this.
One is to derive the heavy-quark theory {\em a priori} in the 
continuum, and then replace the derivatives with difference
operators.\cite{Lepage:1987gg,Lepage:1992tx,%
Eichten:1987xu,Eichten:1990zv,Eichten:1990vp}
These methods are called lattice NRQCD and lattice HQET.%
\footnote{In lattice HQET one treats the $1/m$ corrections as 
insertions.\cite{Eichten:1990vp}
The statistical errors are smaller, however, if one puts the kinetic 
term into the quark propagator.\cite{Lepage:1992ui,Hashimoto:1994sn}
This is still called lattice NRQCD, even when HQET counting is used to 
classify the $1/m$ expansion for heavy-light systems.}
The other is to note\cite{El-Khadra:1997mp,Kronfeld:2000ck} that 
Wilson fermions possess the same heavy-quark symmetries%
\cite{Shifman:1987rj,Isgur:1989vq} as continuum~QCD.
Thus, correlation functions computed with lattice gauge theory can be 
described {\em a posteriori} by HQET (or NRQCD), with a logic and 
structure parallel to the heavy-quark theory for continuum QCD.%
\cite{Kronfeld:2000ck,Harada:2001hl,Harada:2001hh}
Heavy quark symmetry persists for all~$ma$, so the HQET description
can be developed for all $ma$, with minor modifications that are 
discussed below.
These ideas give a systematic procedure for matching lattice gauge 
theory to QCD.
It is sometimes called the non-relativistic interpretation of Wilson 
fermions, and sometimes called the Fermilab method.

For CKM phenomenology, hadrons with one heavy quark are of greatest
interest.
The most important scales are $\Lambda$ and the heavy quark mass~$m$.
In HQET, as used to describe continuum QCD, one separates these two 
scales and, then, treats higher dimensional operators as 
perturbations, to develop a systematic expansion in powers 
of~$\Lambda/m$.
For quarkonium---bound states of a heavy quark and a heavy 
anti-quark---there are three important scales, $m$, $m\nu$, and 
$m\nu^2$, where $\nu$ is the relative velocity between the heavy quark 
and heavy anti-quark.
When $m$ is large enough to probe the Coulombic part of the potential, 
$\nu\sim\alpha_s(m)$ is small.
Each operator in NRQCD must be assigned a power of $\nu$, and the 
effective theory is used to develop an expansion in $\nu$ and 
$\alpha_s$, which are treated as commensurate.

Here we would like to use HQET and NRQCD to understand lattice gauge 
theory with heavy quarks (and moderate lattice spacings).
As long as $m_Q\gg\Lambda_{\rm QCD}$, one can write\cite{Kronfeld:2000ck}
\begin{equation}
	{\cal L}_{\rm lat} \doteq {\cal L}_{\text{HQ}},
	\label{eq:lat=HQ}
\end{equation}
which means that the lattice gauge theory inside the computer can be
described by a heavy quark effective Lagrangian~${\cal L}_{\text{HQ}}$.
The philosophy is in some ways similar, but in other ways different 
from, Eq.~(\ref{eq:lat=Sym}).
The similarity is that we would like to use a continuum field theory 
to describe lattice gauge theory, with an eye to understanding and 
controlling discretization effects.
The difference is that, for a heavy quark, the descriptive field 
theory is built from heavy-quark fields, not from QCD quark fields.
The latter describes both quarks and anti-quarks.\cite{Dirac:1931kp}
The heavy quark field, on the other hand, satisfies a constraint, 
so it corresponds either to quarks, or anti-quarks, but not both.
The arguments supporting Eq.~(\ref{eq:lat=HQ}) are both concrete, 
studying the large mass limit of lattice gauge theory,%
\cite{El-Khadra:1997mp} and abstract, noting (as above) that the 
degrees of freedom and symmetries are right.\cite{Kronfeld:2000ck}

HQET and NRQCD share the same effective Lagrangian, 
\begin{equation}
	{\cal L}_{\text{HQ}} = \sum_n 
		{\cal C}^{\rm lat}_n(m_Q, g^2, m_Qa; \mu/m_Q) {\cal O}_n(\mu),
	\label{eq:HQ}
\end{equation}
where the ${\cal C}_n$ are short-distance coefficients and the 
operators ${\cal O}_n$ encode the long-distance behavior.
The operators do not depend on the short distance scales $1/m_Q$ or~$a$.
It is useful to think of them, as with Symanzik's LE${\cal L}$, 
as being defined with a continuum ultraviolet regulator, 
and some convenient renormalization scheme.
Compared to the HQET/NRQCD description of continuum QCD, the main 
difference is that there are two short distances, $1/m_Q$ and~$a$.
Because the change is at short distance, the short-distance 
coefficients ${\cal C}^{\rm lat}_n$ must be modified: 
they depend on~$m_Qa$, the ratio of short-distance scales.

Let us recall some aspects of heavy-quark theory.
One has
\begin{equation}
	{\cal L}_{\text{HQ}} = {\cal L}^{(0)} + {\cal L}^{(1)} +
		{\cal L}^{(2)} + \cdots.
	\label{eq:LHQET}
\end{equation}
For  HQET ${\cal L}^{(s)}_{\rm HQET}$  contains terms of dimension~$4+s$;
for NRQCD ${\cal L}^{(s)}_{\rm NRQCD}$ contains terms of order~$\nu^{2s+2}$.
In the following, we shall use HQET counting, but the discussion could 
be repeated in NRQCD, with straightforward modifications.
The leading, dimension-four term is 
\begin{equation}
	{\cal L}^{(0)}_{\rm HQET} = \bar{h}_v(iv\cdot D - m_1)h_v,
	\label{eq:L0}
\end{equation}
where $h_v$ is a heavy-quark field satisfying the constraint
\begin{equation}
	i{\kern+0.1em /\kern-0.55em v}    h_v   = h_v, \quad
	\bar{h}_vi{\kern+0.1em /\kern-0.55em v} = \bar{h}_v.
\end{equation}
The choice of the velocity $v$ is somewhat arbitrary.
If $v$ is close to the heavy quark's velocity,
then ${\cal L}^{(0)}$ is a good starting point for the heavy-quark
expansion, which treats the higher-dimension operators as small.
The most practical choice is the containing hadron's velocity.

The mass term in ${\cal L}^{(0)}$ is often omitted.
By heavy-quark symmetry, it has an effect neither on bound-state wave
functions nor, consequently, on matrix elements.
It does affect the mass spectrum, but only additively.
Including the mass obscures the heavy-quark flavor symmetry,
but only slightly.\cite{Kronfeld:2000ck}
For two flavors, let $\theta=(m_{1c}-m_{1b})v\cdot x$; 
then the generators
\begin{equation}
	\tau^1 = \frac{i}{2}\left(
	\begin{array}{cc}
		0 &  e^{i\theta}  \\
		e^{-i\theta} & 0
	\end{array} \right),\quad
	\tau^2 = \frac{i}{2}\left(
	\begin{array}{cc}
		0 & -ie^{i\theta}  \\
		ie^{-i\theta} & 0
	\end{array} \right),\quad
	\tau^3 = \frac{i}{2}\left(
	\begin{array}{cc}
		1 & 0  \\
		0 & -1
	\end{array} \right),
	\label{eq:tau}
\end{equation}
satisfying the SU(2) algebra $[\tau^d,\tau^e]=\varepsilon^{dfe}\tau^f$.
When the mass term is included, higher-dimension operators are
constructed with ${\cal D}^\mu=D^\mu-im_1v^\mu$.\cite{Falk:1992fm}
To describe on-shell matrix elements one may omit operators that 
vanish by the equation of motion, $-iv\cdot{\cal D}h_v=0$, derived 
from Eq.~(\ref{eq:L0}).
Higher-dimension operators are, therefore, constructed 
from ${\cal D}^\mu_\perp=D^\mu_\perp$ and
$[{\cal D}^\mu,{\cal D}^\nu]=[D^\mu,D^\nu]=F^{\mu\nu}$.

The dimension-five interactions are
\begin{equation}
	{\cal L}^{(1)}_{\rm HQET} = {\cal C}^{\text{lat}}_2 {\cal O}_2 +
		{\cal C}^{\text{lat}}_{\cal B} {\cal O}_{\cal B},
	\label{eq:L1}
\end{equation}
where ${\cal C}_2$ and ${\cal C}_{\cal B}$ are short-distance
coefficients, and
\begin{eqnarray}
	{\cal O}_2 & = &
		\bar{h}_vD_\perp^2 h_v, \label{eq:O2} \\
	{\cal O}_{\cal B} & = & 
		\bar{h}_v s_{\alpha\beta}B^{\alpha\beta}h_v,
		\label{eq:OB}
\end{eqnarray}
with $s_{\alpha\beta}=-i\sigma_{\alpha\beta}/2$
and  $B^{\alpha\beta}=\eta^\alpha_\mu\eta^\beta_\nu F^{\mu\nu}$.
In NRQCD, ${\cal O}_2$ scales as $\nu^2$, and must be treating as a 
leading term: 
${\cal L}^{(0)}_{\rm NRQCD}={\cal L}^{(0)}_{\rm HQET}+%
{\cal C}_2 {\cal O}_2$.
In NRQCD, ${\cal O}_{\cal B}$ scales as $\nu^4$, as do several 
operators of dimension six and seven, and this collection of 
operators of order~$\nu^4$ gives the next-to-leading
correction.\cite{Lepage:1992tx}

At dimension six and higher, many operators arise.
The dimension-seven Lagrangian contains the first term to parametrize 
the absence of full rotational symmetry;
\begin{equation}
	{\cal L}^{(3)}_{\rm HQET} = \cdots + {\cal C}_{D^4}^{\text{lat}}
		\bar{h}_v\sum_\mu(D_\perp^\mu)^4h_v.
\end{equation}
It is helpful to think of this operator as appearing in the 
description of continuum QCD too, but with
${\cal C}_{D^4}^{\text{cont}}=0$ enforced by symmetry.

One can also develop an effective field theory description of the 
vector and axial vector currents.\cite{Harada:2001hl,Harada:2001hh}
The details have been worked out for decays of a heavy quark into a 
light quark,\cite{Harada:2001hl} and for decays of a heavy quark into 
another heavy quark,\cite{Harada:2001hh} which is useful for $b\to c$ 
transitions.

We are now in a position to discuss uncertainties in practical 
calculations.
The target is continuum QCD, which can be described along the lines 
given above, with different short-distance coefficients.
The coefficients are
\begin{eqnarray}
	       m_1^{\text{cont}} & = & m, \label{eq:C1cont}  \\
	{\cal C}_2^{\text{cont}} & = & \frac{1}{2m}, \label{eq:C2cont}  \\
	{\cal C}_{\cal B}^{\text{cont}} & = & \frac{z(\mu)}{2m},
	\label{eq:CBcont}
\end{eqnarray}
where $m$ is a renormalized quark mass, and $z$ is a non-trivial 
function of $g^2$ with an anomalous dimension.
At the tree level, $z=1$.
In mass independent renormalization schemes, the renormalized mass 
that appears in Eqs.~(\ref{eq:C1cont})--(\ref{eq:CBcont}) is the 
(perturbative) pole mass.

The description of lattice gauge theory with HQET is useful for
comparing and contrasting the lattice-spacing uncertainties arising
in the various heavy-quark methods.
Since the dependence on $m_Qa$ is isolated into the coefficients, 
heavy-quark lattice artifacts arise only from the mismatch of the 
${\cal C}^{\rm lat}_n$ and their analogs~${\cal C}^{\rm cont}_n$ in 
the description of continuum~QCD.
For brevity, we shall focus on the three most widely used methods, 
namely the extrapolation method, lattice NRQCD, and the Fermilab 
method.
We shall discuss lattice NRQCD first, and then turn to the other two, 
which both use Wilson fermions.

For lattice NRQCD and lattice HQET, the Lagrangian is
\begin{equation}
	{\cal L}_{\text{lat}} = {\cal L}_{\rm gauge}
		- \Psi^\dagger{D_4^+}_{\rm lat}\Psi +
		\sum_n c_n O_n
	\label{eq:latNRQCD}
\end{equation}
where $\Psi$ is a two-component lattice fermion field,
the $O_n$ are discretizations of the higher-dimension~${\cal O}_n$,
and the $c_n$ are free parameters.
The $c_n$ are chosen so that
\begin{eqnarray}
	{\cal C}_2^{\text{lat}} & = & \frac{1}{2m}, \label{eq:C2lat}  \\
	{\cal C}_{\cal B}^{\text{lat}} & = & \frac{z(\mu)}{2m},
	\label{eq:CBlat}
\end{eqnarray}
and so on to the desired order.
Equation~(\ref{eq:C2lat}) identifies what one means by renormalized 
quark mass; it is obtained implicitly, by adjusting a meson's kinetic 
mass to $m_B$ (or $m_\Upsilon$).
Equation~(\ref{eq:CBlat}) is matched in perturbation theory.
(The rest mass is ignored, because it does not affect matrix elements 
or mass splittings.)
Solving for the lattice couplings~$c_n$ one finds, in many cases, 
power-law divergences as $a\to0$.\cite{Lepage:1987gg}
Therefore, lattice NRQCD calculations must keep $a\sim1/m_Q$, and
discretization errors are reduced by keeping more and more terms 
in~${\cal L}_{\text{lat}}$.
One must improve the light quark Lagrangian to the same order.
The restriction on~$a$ is not of much practical importance, because 
the computing challenges discussed in Sec.~\ref{sec:mc} restrict it to 
the same range anyway.

The Fermilab method uses the lattice Lagrangian in Eq.~(\ref{eq:Llat}) 
and adjusts the free parameters of the lattice action according to 
Eqs.~(\ref{eq:C2lat}) and~(\ref{eq:CBlat}).
The solution of these conditions gives the lattice couplings~$c_O$ in
Eq.~(\ref{eq:Llat}) as a function of~$m_Qa$.
In practice, these relations are obtained in perturbation theory.
The key difference to lattice NRQCD is that, as $a\to0$, 
the conventional Symanzik LE${\cal L}$ also applies.
Consequently, the short-distance coefficients of the Fermilab method 
satisfy
\begin{equation}
	\lim_{a\to0} {\cal C}_n^{\text{lat}} = {\cal C}_n^{\text{cont}}.
	\label{eq:C=C}
\end{equation}
Moreover, the corrections to the limiting behavior are related to the 
short-distance coefficients in the Symanzik effective field theory.%
\cite{Harada:2001hl}
The pattern of uncertainties in the Fermilab method depends on~$m_Qa$.
For $m_Qa>1$, discretization effects follow a pattern similar to 
NRQCD, whereas for $m_Qa<1$ the Symanzik theory also is valid. 
On general grounds, one expects the crossover region to be smooth,
and this expectation has been explicitly verified in several cases at 
the one-loop level.\cite{Kuramashi:1998tt,Mertens:1998wx,%
Harada:2001hl,Harada:2001hh}

In the extrapolation method, one (artificially) sets $m_Qa<1$,
and assumes that Symanzik improvement is adequate.
This leaves
\begin{equation}
	{\cal C}_n^{\text{lat}} - {\cal C}_n^{\text{cont}} \sim (m_Qa)^2
	\label{eq:C-Cma}
\end{equation}
for non-perturbative $O(a)$ improvement.
Mass splittings suffer from mismatches of order
$(m_Qa)^n\Lambda/m_Q=(m_Qa)^{n-1}\Lambda a$.
For matrix elements, one must also look at the normalization factor.
Typical recent calculations use a normalization factor based on 
Eq.~(\ref{eq:ZJma}), supplemented with an Ansatz to incorporate full 
tree-level mass dependence from the Fermilab method.%
\cite{Becirevic:1998ua,Bowler:2000tx,Bowler:2000xw,Abada:2000ty}
This leaves an uncertainties of order~$\alpha_s\,(m_Qa)^n$.
As discussed above, this would be fine if $m_Qa$ were small enough.
For the charmed quark, a preliminary study shows that these effects 
are under control for the spectrum, if one takes the continuum 
limit.\cite{Rolf:2001iv}

To get a semi-quantitative feel for the uncertainties, let us consider
a generic quantity $Q$ with heavy quark expansion
\begin{equation}
	Q = Q_0 + \frac{Q_1}{m_Q} + \frac{Q_2}{m_Q^2}.
	\label{eq:Q}
\end{equation}
Table~\ref{tab:hqet} lists the relative uncertainty on $Q$ from each
term.
\begin{table}[tbp]
	\centering
	\caption[tab:hqet]{Parametric uncertainties for heavy quarks.
	For each method (and heavy-light system) the $n$th row gives the 
	relative uncertainty on $Q$ from $Q_n$.
	No estimates are given for the $B$ system in the extrapolation 
	method, because it is not clear how $1/m$~extrapolations  amplify
	the uncertainties.
	Numerical estimates, in percent, are made taking 
	$\alpha_s=0.2$, $\Lambda=500$~MeV, $m_c=1.25$~GeV, $m_b=4$~GeV.
	For illustration, we have taken $a^{-1}=2.5$~GeV.}
	\label{tab:hqet}
	\begin{tabular}{c|ccccc}
	\hline\hline
	method & $O(a)$ & extrap  & 
	\multicolumn{2}{c}{Fermilab} & latNRQCD  \\
	system &  $D$ & $B$  & $D$ &    $B$    &    $B$    \\
	\hline
	$Q_0$ & $\alpha_s (m_ca)^2$ & \quad\quad & $\alpha_s^2$ & 
		$\alpha_s^2$ & $\alpha_s^2$  \\
	   & $~5$ & \quad\quad & $4~$  & $4~$  & $4~$    \\
	$Q_1$ & $m_ca\Lambda a$  &  & $\alpha_s\Lambda a$ & 
		$\alpha_s\Lambda/m_b$ & $\alpha_s\Lambda/m_b$  \\
	   & $10$ &            & $4~$  & $2.5$ & $2.5$  \\
	$Q_2$ & $(\Lambda a)^2$  &  & $\alpha_s\Lambda^2a/m_c$ & 
		$(\Lambda/m_b)^2$ & $(\Lambda/m_b)^2$  \\
	   & $~4$ &            & $1.6$ & $1.6$ & $1.6$  \\
	\hline\hline
	\end{tabular}
\end{table}
The heavy quark expansion is useful here, because in all methods the
physical heavy-quark effects are intertwined with lattice spacing
effects.
We shall assume that Eq.~(\ref{eq:Q}) is adequate for charmed hadrons,
but the conclusions do not really depend on the assumption.
For lattice NRQCD and the Fermilab method, we imagine that most
of the normalization of $Q_0$ is non-perturbative,%
\cite{Hashimoto:2000yp,El-Khadra:2001rv,Hashimoto:2001ds}
but the rest is available at the one-loop level,
and that $Q_1$ is normalized at the tree level.
For the extrapolation method, we imagine full $O(a)$ improvement, 
and neglect the difficulties with $c_A$ and $c_V$ mentioned in
Sec.~\ref{sec:sym}.
Table~\ref{tab:hqet} also includes numerical estimates, made taking
$\alpha_s=0.2$, $\Lambda=500$~MeV, $m_c=1.25$~GeV, $m_b=4$~GeV.

For $B$ physics, lattice NRQCD and the Fermilab method lead to the same 
estimates.
With lattice NRQCD, heavy quark propagators require negligible computing;
with the Fermilab method, they require more computing, but the 
overhead is negligible compared to generating gauge fields.
The main advantage of the Fermilab method is that it has no 
restriction on~$m_Qa$.
No estimate is given for the discretization errors for $B$ physics 
from the extrapolation method.
One has to understand how errors of order $(ma)^n$ propagate through 
the $1/m$ expansion.
One~crude method is to compare different fits; this only tests whether 
the function is smooth in the region where data are available and is, 
thus, an underestimate.

For $D$ physics, the $O(a)$ method and the Fermilab method have 
similar uncertainties, because $m_ca<1$.
As soon as $m_ca<\alpha_s$, non-perturbative improvement becomes 
valuable.
It could be implemented for the Fermilab currents without any 
conceptual difficulty.
Lattice NRQCD is rarely used for charmed hadrons, because
$a^{-1}\sim m_c$ and then perturbation theory in $\alpha_s$, needed
to improve the NRQCD Lagrangian to interesting accuracy, does not
converge well.

To reduce the uncertainty from heavy-quark discretization effects over 
the short term, one probably requires (automated\cite{Nobes:2001tf}) 
perturbation theory.
In lattice NRQCD and the Fermilab method, one more loop would reduce 
each uncertainty by about one~fifth.
Attacking $B$ physics in this way, one could direct increases in 
computing at removing the quenched approximation.
In the extrapolation method, the program to reduce uncertainties is to 
reduce $a$ until there is a window with $m_Qa\ll1$ and 
$\Lambda/m_Q\ll1$ simultaneously.
Unfortunately, this choice postpones unquenched calculations for at 
least another generation.

\section{Light Quark Effects: Chiral Perturbation Theory}
\label{sec:chiral}

Light quarks bring physical scales $m_q\ll\Lambda$ into QCD.
The numerical algorithms for computing the quark propagator slow down 
for light quark masses, as explained in Sec.~\ref{sec:why}.
This problem makes it impractical, as a rule, to carry out numerical
calculations with masses as small as those of the up and down quarks.
To reach the physical region, the Monte Carlo is run at a sequence of 
light quark masses, say in the range $0.2m_s\lesssim m_q\lesssim m_s$, 
and masses and matrix elements are extrapolated to down to $m_d$ 
and~$m_u$.

Because of the artificially large light quark masses, hadrons in the 
computer carry a cloud of light pseudoscalar mesons, all with mass 
squared $m_{PS}^2\lesssim2m_K^2$.
In nature, the cloud contains pions, kaons, and $\eta$ mesons, with a 
wide range of masses, $m_\pi^2\ll m_K^2\approx \case{3}{4}m_\eta^2$.
The relation between the two situations can be understood 
quantitatively, by studying the interaction of the pseudoscalars with 
other hadrons.
If the numerical data are close enough to the chiral limit,
then the machinery of chiral perturbation theory%
\cite{Gasser:1984yg,Gasser:1984gg} ($\chi$PT) can be applied.
Numerical data can be tested against the leading (or next-to-leading
or next-to-next-to-leading) order prediction of $\chi$PT.
If the data verify $\chi$PT, one has a sound guide to extrapolate in
the quark mass.

Chiral perturbation theory is developed by introducing an effective 
Lagrangian built of pion fields and, where appropriate, other hadrons.
In this case, the long distances are the inverse light quark masses, 
whereas the short distance is the QCD scale~$\Lambda$.
The separation scale $\mu$ is, therefore, usually taken to be
around 1~GeV.
Interactions in the chiral Lagrangian can be classified by the number of
derivatives and powers of the quark mass.
The momenta of pseudoscalars and the virtuality of other hadrons is also
assumed to be small, $q^2\sim m_K^2\ll\Lambda^2$.
A concrete scale, less fuzzy than $\Lambda$,
that arises in the expansion is $4\pi f_\pi=1.65$~GeV.

This chapter is not the place for a full review of~$\chi$PT.
Instead, we shall take a specific example---the calculation of the $B$
meson decay constant---to illustrate how $\chi$PT for lattice gauge
theory differs from phenomenological applications to low-energy hadron
physics.
In particular, we will contrast usual $\chi$PT with two situations that
arise in lattice calculations.
The first is quenched $\chi$PT, which is used to describe problems in 
the quenched approximation.
The second is partially quenched $\chi$PT, which applies to computer 
calculations in which the valence quarks have a different mass than 
the sea quarks in loops.
Finally, we discuss how to modify the chiral Lagrangian to take into 
account the fact that lattice QCD usually has less
chiral symmetry than continuum~QCD.

\begin{figure}
	\centering
 	\includegraphics[width=0.8\textwidth]{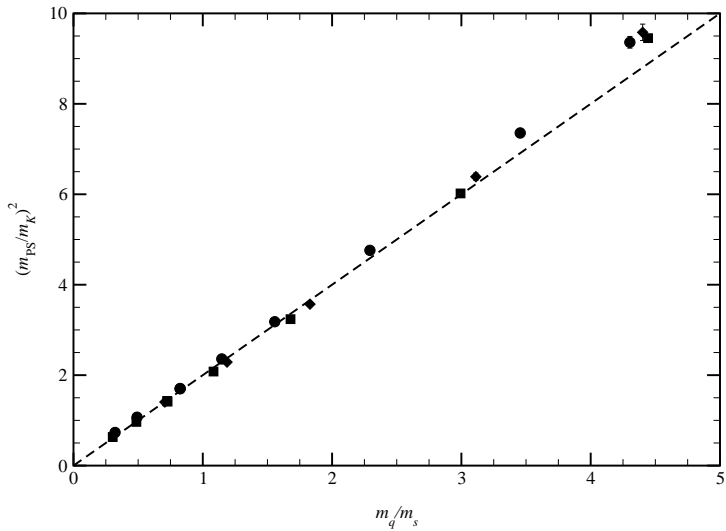}
	\caption[fig:mq]{Plot of the squared pseudoscalar mass~$m^2_{PS}$ 
	vs.\ quark mass~$m_q$, with numerical data in the quenched
	approximation.\cite{Gockeler:1999cy}}
	\label{fig:mq}
\end{figure}

The classic example of a chiral extrapolation is shown in 
Fig.~\ref{fig:mq}.
The pseudoscalar mass is related to the quark mass, 
in leading order, by
\begin{equation}
	m^2_{PS} = (m_q + m_{\bar{q}}) B,
	\label{eq:mPSmq}
\end{equation}
where $B$, from the point of view of chiral perturbation theory, 
is an unknown ``low energy constant.''
From the point of view of lattice QCD, it is calculable.
One simply computes $m_{PS}$ for a variety of (light) quark and 
anti-quark masses and, if the behavior in
Eq.~(\ref{eq:mPSmq}) is verified, fits to obtain~$B$.
Figure~\ref{fig:mq} shows that Eq.~(\ref{eq:mPSmq}) is a good
description out to remarkably large quark masses.
(This may be a fluke of the quenched approximation.)
Indeed, the lattice calculation of $\hat{m}=\case{1}{2}(m_u+m_d)$ is 
nothing but $\hat{m}=m^2_\pi/B$, with $m_\pi=140$~MeV from experiment 
and $B$ from Fig.~\ref{fig:mq}.

The mass of the strange quark is determined in a similar way.
One can set $m_q=m_{\bar{q}}$ and identify $m_s$ as the quark mass 
giving $m_{PS}^2=2m_K^2$.
Alternatively, on can extrapolate in the anti-quark mass, until
$m_{\bar{q}}=\hat{m}\ll m_s$ and identify $m_s$ when
$m_{PS}^2=(m_{K^+}^2+m_{K^0}^2)/2$.
As a rule, in lattice QCD we neglect electromagnetic effects and 
isospin violation---other uncertainties, even statistical errors 
post-extrapolation---are larger.
Note that methods have been developed to include these effects, 
when necessary.\cite{Duncan:1996xy}

Of course, the renormalized quark mass (and, hence, $B$) depends on 
the renormalization scheme.
It is important to take this scheme dependence into account before 
quoting a numerical value.
The two most widely used conventions are the renormalization-group 
invariant mass\cite{Garden:1999fg} and the $\overline{\rm MS}$ 
mass~$\bar{m}_q(2~{\rm GeV})$.
In mass ratios the scheme dependence cancels, and for this reason it 
is convenient to refer to light quark masses in units of the strange 
quark mass, as defined in the previous paragraph.

For other masses and matrix elements it makes more sense to eliminate
the scheme-dependent quark masses in favor of pseudoscalar meson masses.
The meson masses are what one computes numerically, and they are what 
appear in~$\chi$PT.
For example, let is consider $f_{B_q}$, where $q$ is the flavor of the
light quark in the $B$ meson.
We shall neglect $1/m_b$ corrections and write
\begin{equation}
	f_{B_q} = \frac{\Phi}{\sqrt{m_{B_q}}} \left[
		1 + \Delta f_{B_q} \right],
\end{equation}
where $\Phi$ is independent of both heavy and light quark masses, and
$\Delta f_{B_q}$ denotes the (one-loop) contribution of the light meson
cloud.

In QCD, the one-loop correction to the decay constants are
(neglecting isospin breaking)\cite{Boyd:1994pa,Sharpe:1996qp}
\begin{eqnarray}
	\Delta f_{B_s} & = & - \frac{1+3g^2}{(4\pi f)^2} \left[
					 m_K^2    \ln(m_K^2/\mu^2) + 
		\case{1}{3}  m_\eta^2 \ln(m_\eta^2/\mu^2) \right] 
		\nonumber \\ & + &
		c_1(\mu) (m_K^2 + \case{1}{2}m_\pi^2) +
		c_2(\mu) (m_K^2 - \case{1}{2}m_\pi^2) ,
	\label{eq:unquenchedBs} \\
	\Delta f_{B_d} & = & - \frac{1+3g^2}{(4\pi f)^2} \left[
		\case{3}{4}  m_\pi^2  \ln(m_\pi^2/\mu^2) + 
		\case{1}{2}  m_K^2    \ln(m_K^2/\mu^2) + 
		\case{1}{12} m_\eta^2 \ln(m_\eta^2/\mu^2) \right] 
		\nonumber \\ & + &
		c_1(\mu) (m_K^2 + \case{1}{2}m_\pi^2) +
		\case{1}{2}c_2(\mu) m_\pi^2 ,
	\label{eq:unquenchedBd}
\end{eqnarray}
where $f$ and $g$ are (the chiral limit of) the pion decay 
constant and $B$-$B^*$-$\pi$ coupling
The ``low-energy'' constants $c_i$ encode QCD dynamics from distances 
shorter than $\mu^{-1}$, whereas the logarithms are 
long-distance properties constrained by chiral symmetry.
The dependence on $\mu$ cancels in the total.

These formulae illustrate how to derive useful formulae from $\chi$PT.
The light quark masses in numerical lattice calculations are all about
the same size, say $m_q\gtrsim0.2m_s$, so it is not helpful to neglect
$m_\pi^2$ relative to~$m_K^2$.
Because each quark mass is an adjustable parameter in lattice
calculations, general formulae with non-degenerate quark masses are
needed.
If such formulae are available, the combination of lattice calculations
and $\chi$PT is powerful.
Chiral symmetry constrains the coefficient of the logarithmic terms,
tying them to quantities that can be calculated by other means.
This is a key, because it is difficult, from fitting, to distinguish
$m^2\ln m^2$ from a polynomial, unless the range of $m^2$ is very wide.
With the ``chiral logs'' constrained, however, a straightforward
fit the determines the low-energy constants.

\begin{figure}[b]
	\includegraphics[width=0.3\textwidth]{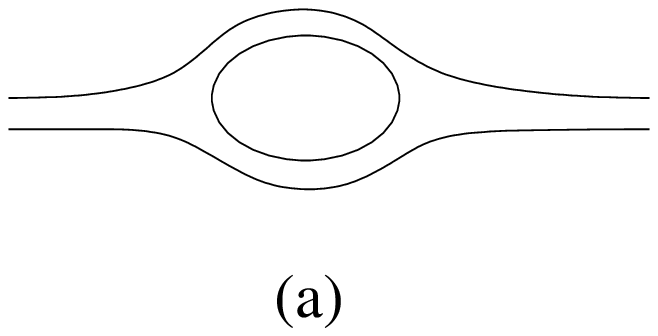}\hspace{.2in}
	\includegraphics[width=0.3\textwidth]{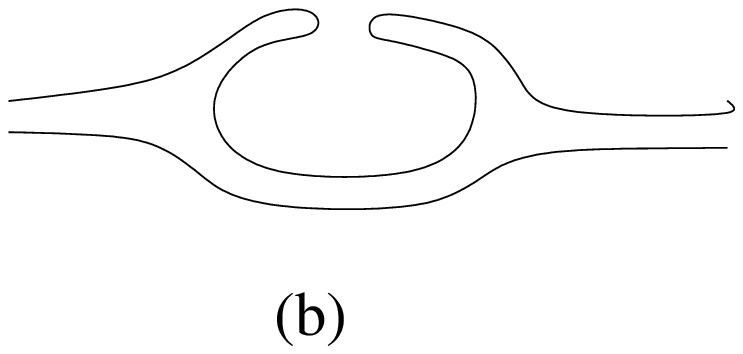}\hspace{.2in}
	\includegraphics[width=0.3\textwidth]{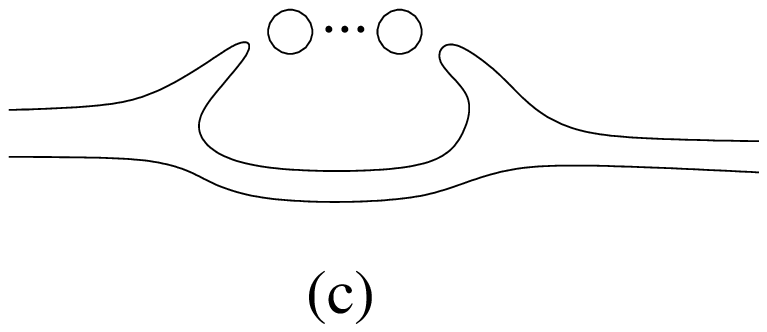}
	\caption[fig:lines]{Quark line configurations that lead to meson
	loops.\cite{Bardeen:2001jm}}
	\label{fig:lines}
\end{figure}

Quenching changes Eqs.~(\ref{eq:unquenchedBs})
and~(\ref{eq:unquenchedBd}) drastically.
Figure~\ref{fig:lines} shows the quark flow of several virtual
processes that take place in a meson.
The quenched approximation includes Fig.~\ref{fig:lines}(b),
but not (a) or~(c).
As a consequence, some pion loops are omitted, and $\eta'$ loops are
mistreated.\cite{Sharpe:1990me}
In a partially quenched calculation, Fig.~\ref{fig:lines}(a) and (c)
are restored, but the masses of quarks in loops are not the same as
those on the valence lines.
In the following, we shall examine how the omission or modification of
these processes changes the quark mass dependence of~$f_B$.

Recall that the quenched (or valence) approximation replaces the 
fermion determinant in Eq.~(\ref{eq:quarkdet}) with~1.
This replacement omits closed fermion loops (and absorbs some of the 
omission into the bare couplings).
Instead of omitting the loops, one could imagine canceling them with 
bosonic loops with the same action.
Written as functional integral\cite{Morel:1987xk}
\begin{equation}
	\frac{\Det({\kern+0.1em /\kern-0.65em D}+m)}
	     {\Det({\kern+0.1em /\kern-0.65em D}+\tilde{m})} = 
		\int {\cal D}q{\cal D}\bar{q} {\cal D}\tilde{q}{\cal D}\bar{\tilde{q}}
		e^{-\bar{q}({\kern+0.1em /\kern-0.65em D}+m)q 
		   -\bar{\tilde{q}}({\kern+0.1em /\kern-0.65em D}+\tilde{m})\tilde{q}}
	\label{eq:ghostDet}
\end{equation}
where $\tilde{q}$ is a {\em bosonic} field, called scalar quarks or
ghost quarks.
We want to understand partially quenched calculations with $n_f$
flavors of sea quarks and $n_v$ flavors of valence quarks.
The mass matrices are
\begin{eqnarray}
	m & = & \diag(m_1,m_1,\ldots,m_{n_f},M_1,M_2,\ldots,M_{n_v}), \\
	\tilde{m} & = & \diag(M_1,M_2,\ldots,M_{n_v}),
\end{eqnarray}
so the ghosts cancel the determinants of the valence quarks.
When studying mesons, one wants $n_v=2$; baryons, $n_v=3$.
The quenched approximation is then the special case~$n_f=0$.

The advantage of Eq.~(\ref{eq:ghostDet}) is that it immediately
suggests the hadronic content of the quenched approximation.%
\cite{Morel:1987xk}
In addition to the usual $\bar{q}q$ mesons, one has bound states of 
ghosts $\bar{\tilde{q}}\tilde{q}$ and bound states of quarks and 
ghosts $\bar{q}\tilde{q}$ and $\bar{\tilde{q}}q$.
States with an odd number of ghosts have negative metric
and some spooky consequences.

If $m$ and $\tilde{m}$ are both small compared to $\Lambda$,
there is an approximate chiral symmetry.
It is not so straightforward to identify the group, because of
differences between bosonic and fermionic integrals, and one must
insist that the determinants cancel.\cite{Damgaard:1998xy,Sharpe:2001fh}
It turns out that, for $n_f\ge1$, that the symmetry group has the
same number of generators as 
\begin{equation}
	G_{\rm PQ} = {\rm SU}(n_f+n_v|n_v) \times {\rm SU}(n_f+n_v|n_v)
		\times{\rm U}(1),
\end{equation}
and the Ward identities derived with this ``intuitively obvious'' group
are the same as for the correct group.
Here ${\rm SU}(n_2|n_1)$ denotes a graded Lie group, a mathematical
beast that also appears in supersymmetry.
The SU groups are spontaneously broken to their diagonal subgroup,
just as in~QCD.

For the quenched approximation, $n_f=0$, the result is slightly
different\cite{Sharpe:2001fh}
\begin{equation}
	G_{\rm Q} = \left[{\rm SU}(n_v|n_v) \times {\rm SU}(n_v|n_v)
		\right] \Join {\rm U}(1),
\end{equation}
where~$\Join$ denotes a {\em semi}-direct product.
The semi-direct product~$\Join$ may be less familiar that the direct
product~$\times$.
Let $A$ and $H$ be subgroups of $G$, and suppose $A$~is normal.
($A$ is normal if, for all $a\in A$, $gag^{-1}\in A$ for all $g\in G$.)
If, furthermore, $A\cap H=\{1\}$, then the semi-direct product
$A\Join H$ consists of all $ah$, $a\in A$, $h\in H$.
This seemingly technical issue has a crucial physical implication:
in quenched $\chi$PT the flavor singlet mesons
($\eta'$ in QCD), as well as singlet ghost-quark and ghost-antighost
mesons, do not decouple.

Consequently, for quenched $\chi$PT, one must construct the effective
Lagrangian for pseudoscalar mesons with a formalism that includes
the $\eta'$.
This can be achieved\cite{Bernard:1992mk} along lines similar to
normal $\chi$PT.\cite{Gasser:1984gg}
The main difference is that the $\eta'$ is not a normal particle.
Its propagator has a double-pole structure, because
Fig.~\ref{fig:lines}(c) is absent.
On the other hand, although it was not obvious until the technical
details were sorted out, the flavor singlets do decouple once there are
closed quark loops to restore
Fig.~\ref{fig:lines}(c).\cite{Sharpe:2001fh}

Let us now quote the results from quenched and partially quenched
$\chi$PT for $\Delta f_B$.
In the quenched approximation\cite{Booth:1994hx,Sharpe:1996qp}
\begin{eqnarray}
	\Delta f_{\bar{b}v}^{\rm Q} & = & - \frac{1}{(4\pi f)^2}
		\left[ \frac{1+3g^2}{6} m_0^2 + \gamma m_{vv}^2
	\right] \ln(m_{vv}^{2}/\mu^2) \nonumber \\
	& + & c^{\rm Q}_0(\mu) m_0^2 + c^{\rm Q}_2(\mu) m_{vv}^2,
	\label{eq:quenchedlog}
\end{eqnarray}
where $\gamma$ is a combination of low energy constants of the quenched
theory, and $m_{vv}$ is the mass of the light pseudoscalar made from
valence quark~$v$.
The constant $m_0^2$ is the residue of the $\eta'$-like double pole.
Finally, $c^{\rm Q}_0$ and $c^{\rm Q}_2$ are counter-terms, whose
dependence on $\mu$ renders $\Delta f_{\bar{b}q}^{\rm Q}$ independent
of~$\mu$.
In general, the low-energy constants of the quenched approximation
cannot be related in any rigorous way to those of~QCD.

The terms proportional to $m_0^2$ diverge in the chiral limit
$m_{vv}^2\to0$.
This behavior shows in detail that the quenched approximation cannot
be forced to account for a wide range of scales.
The common practice to obtain $f_B$ in the quenched approximation does
{\em not} follow from a fit to Eq.~(\ref{eq:quenchedlog}).
Instead one assumes that there is little error from quenching
when $m_v\sim m_s$, and then extrapolates linearly in~$m_{vv}^2$.
Thus, common practice circumvents the clearly unphysical quenched chiral
log, but it must be admitted that the uncertainties
from quenching and chiral logs become intertwined.

In the partially quenched case, the chiral logs behave better.
If all $n_f$ sea quarks have the same mass\cite{Sharpe:1996qp}
\begin{eqnarray}
	\Delta f_{\bar{b}v}^{\rm PQ} & = & - \frac{1+3g^2}{(4\pi f)^2}
		\left[ \frac{n_fm_{vf}^2}{2} \ln(m_{vf}^{2}/\mu^2) +
		\frac{m_{ff}^2-2m_{vv}^2}{2n_f} \ln(m_{vv}^{2}/\mu^2)
		\right] \nonumber \\
	& + & c_1(\mu) m_{ff}^2 + c_2(\mu) m_{vv}^2,
	\label{eq:pqLog}
\end{eqnarray}
which is qualitatively like Eqs.~(\ref{eq:unquenchedBs})
and~(\ref{eq:unquenchedBd}).
A~simple argument relates the partially quenched low-energy constants
to~QCD.\cite{Sharpe:2000bc}
Consider any matrix element to be a function of $n_f+n_v$ sea and
valence masses.
If the valence quark masses are set equal to the sea quark masses,
then one recovers QCD.
Thus, QCD lies on a hyperplane in the space of all masses.
If all masses are sent to zero together, there is a unique chiral limit.
Thus, the low energy constants of the partially quenched theory and QCD
are the same.

The foregoing analysis discussed chiral symmetry as it appears in 
the continuum limit of QCD, and its quenched and partially quenched
theories.
The two most widely used formulations of lattice fermions---Wilson
fermions and staggered fermions---have a smaller symmetry group.
Inspection of Eq.~(\ref{eq:WilsonQuarkAction}) shows that two
sources break $\text{SU}(n_f)\times\text{SU}(n_f)$, namely the mass term
and the part of the difference operator proportional to Dirac matrix~1.
Wilson added the extra terms (or, equivalently, chose such a peculiar
difference operator) to circumvent the so-called doubling problem.

The first action that Wilson tried (and, independently, Jan Smit) was
\begin{eqnarray}
	{\cal L}_{\rm Nq} & = & - m_0 \bar{\psi}(x)\psi(x)
	\label{eq:NaiveQuarkAction} \\
		& - & \frac{1}{2a}\sum_\mu \bar{\psi}(x)
			\gamma_\mu\left[ U_\mu(x)\psi(x+ae_\mu) -
			U^\dagger_\mu(x-ae_\mu)\psi(x-ae_\mu)\right]
	\nonumber
\end{eqnarray}
If $m_0=0$ this action is chirally symmetric.
Unfortunately, the propagator has many poles
\begin{equation}
	S(p) = \frac{a}{i\sum_\mu\gamma_\mu\sin(p_\mu a)},
\end{equation}
where components of the momentum four-vector lie in the range
$-\pi/a<p_\mu\le\pi/a$.
There is a pole at all $2^d$ combinations of $p_\mu a=0,\;\pi$.
These correspond to $2^d$ species of physical particles:
vacuum polarization, for example, gets a factor $2^d$.
Moreover, the axial symmetries are exact: the different species have
a pattern of axial charges such that the anomaly cancels.
To avoid these problems, Wilson decided it was less drastic to break
the axial symmetries explicitly, leaving ${\rm SU}(n_f)$ as the
flavor group.

The staggered formulation of lattice fermions starts with
Eq.~(\ref{eq:NaiveQuarkAction}), but notes that similarity
transformation\cite{Kawamoto:1981hw,Sharatchandra:1981si}
\begin{eqnarray}
	\psi(x) & \mapsto & T(x)\psi(x) \\
	\bar{\psi}(x) & \mapsto & \bar{\psi}(x)T^\dagger(x) \\
	T(x) & = & \gamma_1^{x_1} \gamma_2^{x_2} \gamma_3^{x_3} \gamma_4^{x_4}
\end{eqnarray}
diagonalizes all Dirac matrices.
One then has a sum over four equivalent terms.
One can three of them, leaving $2^4/4=4$ species in the
continuum limit.
The remaining species have a remnant axial U(1)
symmetry.\cite{Sharatchandra:1981si}
On the other hand, the flavor group is smaller than SU(4), and the
notion of flavor is tied up with spacetime symmetries---hence the
name ``staggered.''
The SO(4) Euclidean invariance is broken by the lattice down to the 
semi-direct product $\Gamma_4\Join\text{SW}_4$, where $\Gamma_4$ is 
the Clifford group of $4\times4$ Dirac matrices, and SW$_4$ is symmetry
group of a hypercube.\cite{Kilcup:1986dg,Joos:1987bp}
The factor $\Gamma_4$ is interpreted as a flavor group, and
SW$_4$ as a spacetime symmetry.

Before coming back to the implications of the reduced chiral symmetry,
let us review the nature of the problem.
The Nielsen-Ninomiya theorem states that it is impossible to
maintain chirally symmetry and avoid species doubling with ultralocal
interactions.\cite{Nielsen:1980rz,Friedan:1982nk}
Here {\em ultra}local means that the lattice couplings vanish if the
fields are separated by more than a few lattice spacings.
The~way out, only recently appreciated, is to forgo ultralocality in
favor of locality, which only requires that the couplings fall off
exponentially with separation.
Then, if the lattice Dirac operator~${\kern+0.1em /\kern-0.65em D}$
satisfies\cite{Ginsparg:1981bj}
\begin{equation}
	\gamma_5{\kern+0.1em /\kern-0.65em D} +
	{\kern+0.1em /\kern-0.65em D}\gamma_5 = a
	{\kern+0.1em /\kern-0.65em D}\gamma_5
	{\kern+0.1em /\kern-0.65em D},
	\label{eq:GW}
\end{equation}
which is called the Ginsparg-Wilson relation, then
correlation functions are chirally symmetric.
There are two known solutions to Eq.~(\ref{eq:GW}), one based on
renormalization group ideas related to the original
derivation,\cite{Hasenfratz:1998ri}
and the other related to the Narayanan-Neuberger formulation of chiral
lattice fermions.\cite{Neuberger:1997fp}
Both are local, so they are the basis of a well-behaved field theory,
but not ultralocal, so they avoid the hypothesis of the Nielsen-Ninomiya
theorem.
The solution of Eq.~(\ref{eq:MG=S}) for these formulations is much
more computationally demanding than for Wilson or staggered fermions.
Consequently, they are only beginning to make their way into numerical
calculations.

Table~\ref{tab:chiral} summarizes the flavor symmetry group for lattice
fermions.
\begin{table}[tbp]
	\centering
	\caption[tab:chiral]{Pattern of chiral symmetry breaking for various 
	formulations of lattice fermions.}
	\begin{tabular}{cr@{$\;\to\;$}l}
		\hline\hline
		formulation     & $G$ & $H$ \\
		\hline
		Wilson          & $\text{SU}(n_f)$ & $\text{SU}(n_f)$  \\
		staggered       & $\text{U}(1)\times\Gamma_4$ & $\Gamma_4$  \\
		Ginsparg-Wilson & $\text{SU}(n_f)\times\text{SU}(n_f)$ 
		                & $\text{SU}(n_f)$  \\
		\hline
		continuum QCD   & $\text{SU}(n_f)\times\text{SU}(n_f)$ 
		                & $\text{SU}(n_f)$  \\
		\hline\hline
	\end{tabular}
	\label{tab:chiral}
\end{table}
For staggered fermions, the flavor group comes in a semi-direct
product with the symmetry group of the hypercube, SW$_4$;
for the others it comes in a direct product.
The short-distance coefficients of the chiral Lagrangian must be
modified to depend on $a\Lambda$, the ratio of two short distances.
Because the Symanzik effective field theory shows that the violations
of chiral symmetry come in through higher dimension operators, an
expansion in $a\Lambda$ presents no difficulty, except to introduce
new low-energy constants.
The details depend on how the symmetry is broken.
They have been worked out for
Wilson quarks\cite{Sharpe:1998xm,Rupak:2002sm}
and staggered quarks.\cite{Lee:1999zx}

Now that many groups are acquiring the computer resources needed for
unquenched QCD, the topics discussed in this section will gain in
importance.
Checks of chiral behavior will be important not only for extracting
physics from the numerical calculations, but also for testing the
algorithms.
For example, Sec.~\ref{sec:mc} noted that odd numbers of flavors are
sometimes incorporated into the Monte Carlo by generating a weight
proportional to a fractional power of a determinant.
For example, with staggered fermions one has
$(\det M_{\rm KS})^{n_f/4}$,
because $M_{\rm KS}$ corresponds to four flavors.
It is conjectured,\cite{Bernard:1993sv} that the appropriate graded
group for the continuum limit of this theory is ${\rm SU}(4|4-n_f)$,
like a partially quenched theory with $n_v=4-n_f$.
This conjecture, if true, has important consequences for testing the
physical content of these algorithms, which are otherwise a bit
mysterious.

\section{Finite Spacetime Volume Effects}
\label{sec:volume}

In this section we address systematic effects stemming from the 
infrared cutoff imposed by the finite spacetime volume.
In Euclidean space, the temporal direction starts out on the same
footing as the spatial directions.
Even so, the physical interpretation of finite temporal extent, on
the one hand, and finite spatial volume, on the other, is different.
It turns out that both kinds of effects are valuable, providing tools
to extend the range of problems to which Euclidean, numerical lattice
QCD can be applied.

In Sec.~\ref{subsec:T}, we shall address the implications of the finite
temporal extent, and the relation to QCD thermodynamics.
In Sec.~\ref{subsec:V}, we turn to the effects of the finite spatial
volume.
Several issues arise here.
Generic corrections to masses and $1\to0$ and $1\to1$ matrix elements
are suppressed by a factor $e^{-\mu L}$, where $\mu$ is a mass related
to the mass of the lightest hadron (assumed massive).
The effects are larger, suppressed by powers of $L$ instead of an
exponential, when pseudo-Goldstone bosons satisfy $Lm_\pi\lesssim1$.
The most intriguing effect of finite volume is on scattering states,
whose allowed energies are connected in a model-independent way to
final-state phase shifts.

\subsection[Finite temporal extent]{Finite %
temporal extent (non-zero temperature)}
\label{subsec:T}

Recall that numerical lattice calculations are formulated in Euclidean 
space, namely with imaginary time.
In the computer, the extent of the imaginary time is finite,
$L_4=N_4a$.
In Sec.~\ref{sec:mc}, starting with Eq.~(\ref{eq:C2}), we blithely
assumed a correspondence between correlation functions and vacuum
expectation values of time-ordered products, viz.,
\begin{eqnarray}
	\frac{1}{Z} \int \prod_{x,\mu} dU_\mu(x) 
		\prod_x d\psi(x) d\bar{\psi}(x) &
		O_1(t_1)O_2(t_2)\cdots O_n(t_n) & e^{-S} \to \nonumber \\  
		\bra{0}T& O_1(t_1)O_2(t_2)\cdots O_n(t_n) & \ket{0}.
	\label{eq:Cn}
\end{eqnarray}
When $L_4=N_4a$ is finite, however, this correspondence is not exact.

The essential features are easily seen for continuous time.
Then the Euclidean action is simply related to a Lagrangian 
(of a system with several degrees of freedom at each lattice site).
From this Lagrangian one can straightforwardly derive a Hamiltonian
and set up the equivalence between the path-integral and canonical 
formulations of quantum mechanics.
To be explicit, consider the case with only one operator $O(t)$ in 
Eq.~(\ref{eq:Cn}). 
Also let us start with fixed boundary conditions, which means that 
fields at $x_4=0$ and~$x_4=L_4$ are not integrated over. 
Then,
\begin{eqnarray}
	\langle O(t) \rangle & = & \frac{1}{Z} 
		\bra{f} e^{-\hat{H}(L_4-t)} \hat{O} e^{-\hat{H}t}\ket{i}, 
	\label{eq:<O>} \\
	Z & = & \bra{f} e^{-\hat{H}L_4} \ket{i},
\end{eqnarray}
where $\ket{i}$ and $\bra{f}$ are the initial and final states implied 
by the boundary conditions, $\hat{H}$~is the Hamiltonian, and
$\hat{O}$~is the Hilbert-space operator corresponding to the function 
(of quark and gluon fields)~$O$ on the left-hand side.
Inserting complete sets of eigenstates of $\hat{H}$ into 
Eq.~(\ref{eq:<O>}) one sees
\begin{equation}
	\langle O(t) \rangle = \frac{\sum_{mn}\bracket{m}{\hat{O}}{n}%
		e^{-E_m(L_4-t)} e^{-E_nt}\braket{f}{m} \braket{n}{i}}{%
		\sum_n e^{-E_nL_4} \braket{f}{n} \braket{n}{i}}, 
	\label{eq:<m|O|n>}
\end{equation}
where the sums range over all eigenstates of the Hamiltonian, 
and $E_n$ is the energy of state~$\ket{n}$.
For large $L_4$ the ground state---or vacuum---dominates the sum in the 
denominator:
\begin{equation}
	Z\to e^{-E_0L_4} \braket{f}{0} \braket{0}{i}.
\end{equation}
Similarly, if $t$ and $L_4-t$ are large
\begin{equation}
	\langle O(t) \rangle = \bracket{0}{\hat{O}}{0} +
		O\left(e^{-(E_1-E_0)\min(L_4-t,t)}\right), 
	\label{eq:<O>contaminated}
\end{equation}
with the vacuum energy and overlap factors canceling out of the 
ratio in Eq.~(\ref{eq:<m|O|n>}).
One can repeat this analysis for operators at several times and, 
in this way, derive Eq.~(\ref{eq:Cn}).

Although fixed boundary conditions are sometimes useful,%
\cite{Mutter:1981fa,Guagnelli:1999zf} the contamination from excited
states is unnecessarily large.
This is especially so in two- and three-point functions,
where one wants several time separations to be large, 
cf.\ Eqs.~(\ref{eq:C2})--(\ref{eq:CQexp}).
To reduce the contamination term in Eq.~(\ref{eq:<O>contaminated}) it 
is helpful to choose boundary conditions so that
$\ket{f}=e^{-i\phi_i}\ket{i}$, and then to sum over~$i$.
Then Eq.~(\ref{eq:<m|O|n>}) becomes
\begin{equation}
	\langle O(t) \rangle = \frac{%
		\sum_{i}e^{i\phi_i}\bracket{i}{\hat{O}}{i} e^{-E_iL_4} }{%
		\sum_{i}e^{i\phi_i} e^{-E_iL_4}}, 
	\label{eq:<n|O|n>}
\end{equation}
and for large $L_4$ but {\em any}~$t$
\begin{equation}
	\langle O(t) \rangle = \bracket{0}{\hat{O}}{0} +
		O\left(e^{-(E_1-E_0)L_4}\right).
	\label{eq:L4corr}
\end{equation}
One has $\ket{f}=e^{-i\phi_i}\ket{i}$ if the fields are periodic
up to phases, rotations in flavor, and discrete symmetry operations.
For example, to set $\ket{f}=\ket{i}$, with no non-trivial phase,
one chooses periodic boundary conditions for bosonic variables
and {\em anti}-periodic for fermionic variables:
\begin{equation}
	A_\mu(L_4,\bbox{x})= + A_\mu(0,\bbox{x}), \quad 
	\psi(L_4,\bbox{x}) = -  \psi(0,\bbox{x}).
	\label{eq:pbcT}
\end{equation}
The minus sign for fermions follows from Fermi-Dirac statistics.
The sum over initial state $\ket{i}$ is achieved by integrating over 
the fields at $x_4=0$.

Let us focus for a while on the boundary conditions in
Eq.~(\ref{eq:pbcT}).
Then one can re-write Eq.~(\ref{eq:<n|O|n>}) as
\begin{equation}
	\langle O(t) \rangle =
		\frac{\Tr\hat{O}e^{-\hat{H}L_4}}{\Tr e^{-\hat{H}L_4}}.
	\label{eq:thermal}
\end{equation}
Thus, the Euclidean functional integrals give thermal expectation
values for a quantum system with temperature
\begin{equation}
	T = (k_BL_4)^{-1},
	\label{eq:T}
\end{equation}
where $k_B$ is Boltzmann's constant.
Calculations with non-zero temperatures $T\sim\Lambda/k_B$ are of 
considerable interest in nuclear physics and astrophysics.
When the temperature (and also the chemical potential for baryons) is 
in this range, one expects a phase transition.
The hot, dense phase is called the quark-gluon plasma, and it is
thought to exist in heavy-ion collisions\cite{Satz:2000bn} and in
quark stars,\cite{Fraga:2001id} which are similar to neutron stars,
but with large enough gravitational attraction to crush the neutrons
into the quark-gluon plasma.

Lattice calculations have been used to compute the critical temperature
of the phase transition.
In the pure gauge theory, this was one of the
first large-scale calculations without guidance
from experiment.\cite{Gottlieb:1985ug,Christ:1986wx}
The critical temperature for QCD with 2 and 4 flavors has also been
computed.\cite{Gavai:1989pr,Brown:1990ev,Fukugita:1990vu}
Over the last several years, many issues in QCD thermodynamics have
been studied, such as the equation of state of the quark-gluon
plasma.\cite{Boyd:1995zg}
For further reading on this rich application of lattice QCD,
there are several recent reviews.\cite{Laermann:1999qd}

For applications of lattice QCD motivated by particle physics, 
we are interested in zero temperature.
In practice, $L_4$ is large but not infinite, 
so $T$ is small but non-zero.
From Eq.~(\ref{eq:L4corr}), one sees that the corrections are small: 
the energy gap between a state and the vacuum is nothing but the 
energy of a particle, as conventionally defined.
In quenched calculations, $E_1-E_0$ is the glueball mass.%
\footnote{One can consider quenched QCD to contain $\bar{q}q$ mesons
(and $qqq$ baryons) in the Hilbert space.
But then one must also include $\bar{q}\tilde{q}$ mesons, etc.
The contributions of genuine hadrons and the ghosts cancel
in Eq.~(\ref{eq:<n|O|n>}).}
With dynamical light quarks, the energy gap will be the ``pion'' mass, 
that is, the mass of the lightest pseudoscalar~$m_{PS}$.
The exponential suppression means that to reduce the thermal correction 
below 0.1\% one needs $m_{PS}L_4>7$ or $L_4\gtrsim4$~fm for 
$m_{PS}\sim350$~MeV.  

The discussion of Eqs.~(\ref{eq:<O>})--(\ref{eq:thermal}) assumed 
continuous (Euclidean) time.
For discrete time, the analysis stays the same, but the relation 
between the Lagrangian and the Hamiltonian is a little different.
The essential idea is as follows.
There is, perhaps not surprisingly, a natural time-evolution operator, 
called the transfer operator (or, more often, the transfer matrix).%
\cite{Wilson:1974sk,Wilson:1975hf,%
Creutz:1976ch,Luscher:1976ms,Osterwalder:1977pc}
It propagates states in Hilbert space forward one unit of Euclidean 
time:
\begin{equation}
	\ket{\Omega(x_4+a)} = \hat{\Bbb T} \ket{\Omega(x_4)},
	\label{eq:Top}
\end{equation}
where $\ket{\Omega}$ is any state in the Hilbert space.
For details on the relationship between the lattice action and the
transfer matrix, and how to define the Hilbert space,
the reader should consult one of the 
textbooks.\cite{Creutz:mg,Rothe:kp,Montvay:cy} 
The Hamiltonian~$\hat{H}$ is then defined by solving
\begin{equation}
	\hat{\transfer} = \exp(-a\hat{H}).
	\label{eq:H}
\end{equation}
Indeed, all the masses considered in Sec.~\ref{sec:mc} are eigenvalues 
of this Hamiltonian.
The maximal eigenvalue ${\transfer}_0$ is simply $e^{-aE_0}$,
where $E_0$ is interpreted as the vacuum energy.
Equation~(\ref{eq:H}) provides an adequate and useful definition of 
$\hat{H}$, if $\hat{\transfer}$ is positive and has several 
eigenvalues close to the vacuum.
Then $-a^{-1}\ln(\transfer_n/\transfer_0)$ are just the particle masses
and energies.
The transfer matrix is indeed positive for simple actions, like the 
Wilson action.
For improved actions there can be non-positive eigenvalues;
unless the action is pathological (and, thus, not an improvement)
such eigenstates have energies near the cutoff and are, therefore, not 
physical.

\subsection{Finite spatial volume}
\label{subsec:V}

This subsection discusses corrections from the finite spatial volume.
There are two classes of effects:
polarization effects and scattering effects.
The former arise because interacting particles are surrounded by a
cloud of virtual particles.
In a periodic volume, these virtual particles can propagate ``around the
world.''
Except when the cloud contains relatively light particles with
$Lm\lesssim1$, these are exponentially suppressed.
The scattering effects are much more interesting.
The boundary influences the energies of two-particle states in a way
that is connected to the phase shifts.
As a consequence, calculations of the $L$ dependence of these energies
yields valuable information, such as scattering lengths, resonance
widths, and so~on.
We shall aim for a somewhat simplified explanation.
A~more mathematical review is in van~Baal's
chapter.\cite{vanBaal:2000zc}

We consider the size~$L$ of the box to be large enough that the 
finite volume does not alter the structure of the hadrons, namely 
$L\Lambda\gg1$.
Then finite-volume effects are at long distances (by definition), so 
the appropriate degrees of freedom, in QCD, are hadrons.
As long as the ``pion'' is not too light, namely $Lm_{PS}\gg1$,
the correct effective quantum field theory includes a massive field for
each hadron.
Unlike the effective theories of Secs.~\ref{sec:sym}--\ref{sec:chiral},
there is no split here into a leading term and small corrections.
Instead, the hadronic field theory provides a general 
parametrization that is consistent with analyticity, unitarity, and 
symmetries.
The utility of such a theory for describing finite-volume effects was
first pointed out by L\"uscher,\cite{Luscher:1985dn,Luscher:1986pf}
and most of the results sketched below are due to him.%
\cite{Luscher:1985dn,Luscher:1986pf,Luscher:1991ux,Luscher:1991cf}

For the polarization effects, L\"uscher used an all-orders skeleton
expansion for self-energies and vertex functions.
The main features can be appreciated at the one-loop level.
For simplicity, let us consider only two particles, a ``nucleon'' and 
a ``pion'', with effective Lagrangian
\begin{equation}
	{\cal L} = -\bar{N}\left({\kern+0.1em /\kern-0.55em \partial} 
		+ m_{0N}\right)N 
		- \case{1}{2}\left(\partial_\mu\pi\right)^2
		- \case{1}{2}m_\pi^2 \pi^2
		+ y \pi\bar{N}\gamma_5N.
	\label{eq:piN}
\end{equation}
The Yukawa interaction leads to the nucleon self-energy
\begin{equation}
	\Sigma_L(p) = y^2\int \frac{dk_4}{2\pi}
		\frac{1}{L^3}\sum_{\bbox{\nu}} 
		\frac{i({\kern+0.1em /\kern-0.55em p}
			  + {\kern+0.1em /\kern-0.55em k})+m_{0N}}%
			{(k^2+m_\pi^2)[(p+k)^2+m_{0N}^2]},
	\label{eq:Nself}
\end{equation}
where the spatial momentum is discrete in a finite box.
For periodic boundary conditions, $\pi(x+Le_i)=\pi(x)$, components of
the spatial momentum satisfy
\begin{equation}
	e^{ik_iL} = 1 \Rightarrow
	k_i = \frac{2\pi}{L}\nu_i,
	\label{eq:knu}
\end{equation}
where $\nu_i$ is an integer.
Our aim is to exhibit the difference between the mode sum in 
Eq.~(\ref{eq:Nself}) and an integral over all~$\bbox{k}$.

The first step is to use an exponential representation of the 
propagators.
Then the self energy becomes
\begin{eqnarray}
	\Sigma_L(p) & = & y^2 \int_0^\infty d\rho \int_0^1 dx
		\int \frac{dk_4}{2\pi} \frac{1}{L^3}\sum_{\bbox{\nu}} \left[
		i({\kern+0.1em /\kern-0.55em p} + 
		  {\kern+0.1em /\kern-0.55em k})+m_{0N} \right] \nonumber \\
		  & & \times \rho\,e^{-\rho[k^2 + 2xp\cdot k]} 
		  e^{-\rho[(1-x)m_\pi^2 + x(p^2+m_{0N}^2)]}.
	\label{eq:Schwing}
\end{eqnarray}
Now the three sums over spatial momenta are disentangled to the form
\begin{equation}
	{\cal I}(r,s) = \frac{1}{L} \sum_{\nu_i=-\infty}^{\nu_i=\infty}
		e^{-r(\nu_i^2+2s\nu_i)},
	\label{eq:Ii}
\end{equation}
where, in our case, $r=\rho(2\pi/L)^2$ 
and~$s=s_i\equiv xp_iL/2\pi$ for each~$i$.
The terms with $k_i\gamma_i$ in Eq.~(\ref{eq:Schwing}) require 
\begin{equation}
	\frac{1}{L} \sum_{\nu=-\infty}^{\nu=\infty} \nu\,
		e^{-r(\nu^2+2s\nu)} = - \frac{1}{2r} 
		\left. \frac{\partial{\cal I}}{\partial s}\right|_{s=s_i}.
\end{equation}
The sum ${\cal I}$ can be rearranged with the help of the Poisson 
resummation formula
\begin{equation}
	{\cal I}(\rho(2\pi/L)^2,s_i) = \frac{1}{\sqrt{4\pi}}
		\frac{e^{\rho x^2p_i^2}}{\sqrt{\rho}} 
		\sum_{n  =-\infty}^{+\infty} \cos(nxp_iL)
		\exp\left(\frac{-n^2L^2}{4\rho}\right).
	\label{eq:Poisson}
\end{equation}
When $L$ is large, the term with $n=0$ dominates;
it is exactly the same as the the integral over all~$k_i$.
The leading finite-volume correction comes from the terms with $n=\pm1$.
Thus,
\begin{eqnarray}
	{\cal I}(\rho(2\pi/L)^2,s_i) & = & \frac{1}{\sqrt{4\pi}}
		\frac{e^{\rho x^2p_i^2}}{\sqrt{\rho}} 
		\left(1 + 2\cos(xp_iL) e^{-L^2/4\rho} \right) \\
	 	& = & \int\frac{dk}{2\pi} e^{-\rho(k^2+2xp_ik)} 
	 	\left(1 + 2\cos(xp_iL) e^{-L^2/4\rho} \right)
\end{eqnarray}
up to terms suppressed by $e^{-L^2/\rho}$.

To derive the finite-volume correction to the nucleon mass, it is 
enough to set $p$ on shell: $p_4=im_{0N}\gamma_4$, $\bbox{p}=\bbox{0}$.
Then,
\begin{eqnarray}
	\Sigma_L & = & \frac{y^2m_{0N}}{(4\pi)^{d/2}} 
		\int_0^\infty d\rho \int_0^1 dx\, x\, \rho^{1-d/2}\, 
		e^{-\rho[(1-x)m_\pi^2+x^2m_{0N}^2]} \nonumber \\ & & \times\;
		\left(1 + 2 e^{-L^2/4\rho} \right)^{d-1},
	\label{eq:Sigma}
\end{eqnarray}
in $d$ dimensional spacetime.
Eq.~(\ref{eq:Sigma}) is remarkable for several reasons.
It naturally lends itself to dimensional regularization of ultraviolet
divergences, which, when $d\ge4$, arise from the lower limit of the 
integration over~$\rho$.
But in the finite-volume corrections the exponential factors
$e^{-n^2L^2/4\rho}$ suppress this region, faster than any power
of $\rho$.
Therefore, renormalization of ultraviolet divergences is carried out the
same way as in infinite volume.
In Eq.~(\ref{eq:Sigma}), the divergence is removed by mass
renormalization.
The asymptotic finite-volume correction can be evaluated using the
method of steepest descent to carry out the integration over~$x$
and~$\rho$.
Assuming $m_\pi^2<2m_N^2$,
\begin{equation}
	m_N(L) - m_N(\infty) = \Sigma_\infty - \Sigma_L = 
		- \frac{3y^2}{8\pi} \frac{m_\pi^2}{m_N^2} \frac{e^{-L\mu}}{L}
	\label{eq:Deltam}
\end{equation}
where $\mu^2=m_\pi^2(1-m_\pi^2/4m_N^2)$.
The finite volume mass shift is less then 0.1\% if $L\mu>7$.

The $L$ dependence holds for any stable particle and to all orders
in the interaction.\cite{Luscher:1985dn}
(Resonances, such as the $\rho$ meson and the $\Delta$ baryon, are
another matter; see below.)
If both particles in the self energy have the same mass~$m$,
then $\mu=\sqrt{3}m/2$.
In the case of the physical pion and nucleon, $m_\pi\ll2m_N$, so 
$\mu\approx m_\pi$.
Even if the lightest pseudoscalar has $m_{PS}=400$~MeV, which
corresponds to a light quark mass~$m_s/3$, then $\mu\approx m_{PS}$.

For matrix elements with at most one hadron in the final state,
one can carry out a similar analysis.
The basic structure remains the same: use an exponential representation 
of the propagators and then evaluate the mode sums with the Poisson 
formula.
In this way, it follows that the finite volume corrections are
exponentially suppressed (as long as $Lm\gg1$ for all particles).

Equation~(\ref{eq:Deltam}) holds only if $Lm_\pi$ is large.
When $Lm_\pi\lesssim1$, the integration over $x$ in Eq.~(\ref{eq:Sigma})
is qualitatively different.
For $m_\pi=0$, the $x$ integration is elementary, leading to
\begin{equation}
	m_N(L) - m_N(\infty) = - \frac{3y^2}{4\pi^2}
		\frac{1}{L^2m_N} \quad ({\rm for}~m_\pi=0).
	\label{eq:DeltamChiral}
\end{equation}
Thus, as one would expect, finite-volume corrections are larger when 
there is a cloud of massless particles.
Equation~(\ref{eq:DeltamChiral}) is a simple example of finite-volume
chiral perturbation theory, which can be used rather generally for both
finite volume and non-zero temperature.%
\cite{Gasser:1987zq,Hasenfratz:1989pk,Hansen:1990un}
Section~\ref{sec:chiral} noted that the logarithms of chiral loops are
difficult to demonstrate in numerical data.
That is unfortunate, because chiral behavior is a good
{\em a posteriori} check of algorithms for full~QCD;
indeed, the check may be easier to understand than the algorithms
themselves.
It may, therefore, prove worthwhile to hold $L$ artificially small, and
test for finite-volume Goldstone-boson effects.

Before turning to finite-volume effects in scattering states, let
us consider how to extract hadron masses and matrix elements with
physical light quark masses and infinite volume, from numerical data
with practical light quark masses and finite volume.
In practice, the lightest pseudoscalar is indeed massive, because of 
the chiral slowing down of the numerical algorithms, Eq.~(\ref{eq:mqslow}).
For the sake of illustration, let us assume the lightest pseudoscalar 
has a mass $m_{PS}>400$~MeV.
A~tractable situation has $L=2$~fm, $L_4=2L$.
The exponential factor $e^{-m_{PS}L}$ then falls below~2\% 
(and $e^{-m_{PS}L_4}$ below 0.05\%).
Up and down quarks are reached by fitting data with $m_s/3<m_q<m_s$
to the guide given by infinite-volume chiral perturbation theory.
Although the finite-volume corrections from Eq.~(\ref{eq:Deltam})
are present in the data, they induce only a small bias in the fit.
The fit parameters inherit the bias, but when reconstituting the desired
hadronic property from the fit parameters and physical quark masses,
the finite-volume effect remains under control.


Finite-volume effects are most interesting for scattering states,
including resonances, such as the $\rho$ meson and the $\Delta$ nucleon.
Using models Wiese\cite{Wiese:1988qy} and DeGrand\cite{DeGrand:1990ip}
showed that it would be difficult to determine resonance masses in a
finite volume.
The main difficulty is that a resonance has the same quantum
numbers as multi-particle states.
Intuitively, the latter depend strongly on~$L$, because the natural 
unit of relative momenta is $2\pi/L$, but a resonance's mass should be 
nearly independent of~$L$.
Hence, there are many would-be level crossings.
Of course, levels cannot cross as a parameter (here $L$) varies 
smoothly; instead the states interact and the levels push each other 
apart.
It is plausible that this phenomenon is sensitive to details of the 
dynamics, but details are not immediate.

A~genuine breakthrough came in L\"uscher's 1991 paper,%
\cite{Luscher:1991ux} which derived a non-perturbative relation
between two-particle energies and phase shifts.
In a finite volume, the energy spectrum is discrete.
The main insight is that the phase shift is generated at relatively 
short range, whereas the boundary conditions that render a discrete 
spectrum are imposed at very long distances.
Therefore, the two aspects of the problem can be attacked separately.
Furthermore, it is clear from this consideration that it is the 
infinite volume phase shift that should arise.

The relation between the spectrum and phase shifts has been proven for 
elastic scattering.
In the inelastic region, conceptual problems with the definition of 
finite-volume $n$-particle states ($n>2$) have not been worked out.
For pion-pion scattering, the elastic region is $2m_\pi<E<4m_\pi$;
for pion-nucleon scattering, $m_N+m_\pi<E<m_N+3m_\pi$.
Particularly in numerical lattice QCD, where the pion is artificially 
heavy, these limits apply to the $\rho$ and~$\Delta$.
Indeed, in many examples we suppose that $m_{PS}\gtrsim300$~GeV;
for $\rho$ and~$\Delta$ resonance properties, one should reduce the 
pseudoscalar mass further.

In infinite volume, the scattering phase shift arises in connecting 
the short- and long-range parts of the two-body wave function.
In QCD long range means distances greater than $r_0\sim1$--2~fm.
Thus, the phase shift encodes the part of the confining interactions
of quarks and gluons that is ``remembered'' after hadrons have scattered.
At very large separation, the only boundary condition on the two-body 
wave function is that its scattered component fall off like a 
spherical wave.
All energies are possible and chosen by the initial conditions.
Thus the scattering amplitude (which depends on the phase shifts) is 
the most interesting object, and experiments measure cross sections.

In a periodic finite volume, several features are radically different.
Recall that one should take the time extent $L_4$ is at least as big as 
$L$, to set the temperature to zero.
In that case, the two ``scattering'' particles either never separate 
very far (if their relative momentum is small) or encounter each other 
more than once (if large).
So, a straightforward scattering problem does not arise;
scattering amplitudes and cross sections are not natural observables.
On the other hand, the two-body wave function has to be periodic,
\begin{equation}
	\psi_k(\bbox{r}+\hat{\bbox{e}}_iL) = \psi_k(\bbox{r}),
	\label{eq:2bodyPBC}
\end{equation}
where $\bbox{r}$ is the relative separation of the two particles, and
the label $k$ corresponds to the relative momentum.
The energy of the two-body state is
\begin{equation}
	E = 2\sqrt{m^2 + k^2},
	\label{eq:Ek}
\end{equation}
when both particles are assumed to have the same mass~$m$.
For fixed~$L$, Eq.~(\ref{eq:2bodyPBC}) allows only certain values 
of~$k$.
Because the boundary condition is a long-distance effect, the only 
aspect of the interaction that can enter is the set of phase shifts.
The allowed values can be compactly (but impenetrably) summarized as 
the solutions to\cite{Luscher:1991ux}
\begin{equation}
	\det\left[e^{2i\delta(k)}-U(kL/2\pi)\right] = 0,
	\label{eq:phasedet}
\end{equation}
where the matrix indices are azimuthal and magnetic angular momentum 
quantum numbers $l$ and $m$, $U_{lm,l'm'}$.
The phase shift matrix is
\begin{equation}
	\left[e^{2i\delta(k)}\right]_{lm,l'm'} = e^{2i\delta_l(k)} \delta_{ll'},
	\label{eq:deltaMatrix}
\end{equation}
where $\delta_l(k)$ is the phase shift in the $l$-th partial wave, and 
(in an unfortunate clash of notation) $\delta_{ll'}$ is the Kronecker 
delta.

The most important feature of Eq.~(\ref{eq:phasedet}) is
that $U(q)$ is built from completely kinematical,
$L$ independent functions.\cite{Luscher:1991ux}
Moreover, $U$ splits into blocks along the diagonal, with a block for 
each irreducible representation of the cubic group.
Finally, if, in each block,
higher partial waves have small phase shifts,
then Eq.~(\ref{eq:phasedet}) reduces to a simple equation.

To illustrate this structure, let us consider the case of the $\rho$ 
meson, considering the practical situation $m_\rho<4m_\pi$.
First let us neglect all phase shifts with $l>4$.
Then Eq.~(\ref{eq:phasedet}) becomes
\begin{equation}
	\left\|\begin{array}{cc}
		e^{2i\delta_1(k)} - u_{11} & - u_{13} \\
		- u_{31} & e^{2i\delta_3(k)} - u_{33} \end{array}\right\| = 0.
	\label{eq:rhodet}
\end{equation}
This is tractable, because the information needed to get~$u_{ij}$ has 
been tabulated.\cite{Luscher:1991ux}
It is even simpler if one can neglect $\delta_3$.\cite{Luscher:1991cf}
Then
\begin{equation}
	\delta_1(k) = n\pi - \phi(kL/2\pi),
	\label{eq:delta}
\end{equation}
where
\begin{eqnarray}
	\tan\phi(q) & = & - \frac{\pi^{3/2}q^2}{{\cal Z}_{00}(1;q^2)} \\
	\label{eq:tanphi}
	{\cal Z}_{00}(s;q^2) & = & \frac{1}{\sqrt{4\pi}}
		\sum_{\bbox{\nu}} \frac{1}{(\bbox{\nu}^2-q^2)^s} ,
\end{eqnarray}
with the sum over three-vectors of integers.
The zeta function ${\cal Z}_{00}$ (and several cousins ${\cal Z}_{lm}$) 
are defined by the sum for $\real 2s>l+3$ and by analytic 
continuation for $s=1$.
$\phi(q)$ is a monotonically increasing function of $q$;
as one pieces together the phase shift, the integer $n$ is chosen so 
that $\delta_1(k)$ is smooth.

To see how this works, let us outline how to obtain the $\rho$ mass 
and width.\cite{Luscher:1991cf}
The first step is to compute, via Monte Carlo, the $I=1$ two-pion
energies~$E$, as a function of~$L$.
Via Eq.~(\ref{eq:Ek}) each energy yields a momentum
$k=\case{1}{2}\sqrt{E^2-4m_\pi^2}$, and then the phase shift~$\delta_1(k)$
is obtained from Eq.~(\ref{eq:delta}).
The $\rho$ resonance appears at momentum~$k_\rho$, such that
$\delta_1(k_\rho)=\pi/2$.
Then, by studying the $L$ (and hence $k$ dependence) around the
resonance, the effective range formula
\begin{equation}
	\frac{k^3}{E}\cot \delta_1(k) =
		\frac{4k_\rho^3(k_\rho^2-k^2)}{m_\rho^2\Gamma_\rho}
\end{equation}
can be used to determine the width.
The resulting $(m_\rho,\Gamma_\rho)$ would correspond to an 
unphysical case, with too large quark masses.
From chiral perturbation theory, we expect $m_\rho$ to depend mildly 
on quark masses.
Similarly, if the decay is modeled with the interaction
\begin{equation}
	{\cal L}_{\rho\pi\pi} = g_{\rho\pi\pi}
		\varepsilon_{abc}\rho_\mu^a\pi^b\partial^\mu\pi^c,
\end{equation}
then $\chi$PT suggests that $g_{\rho\pi\pi}$ depends only mildly on
the mass.
The width derived from ${\cal L}_{\rho\pi\pi}$ is
\begin{equation}
	\Gamma_\rho = \frac{g_{\rho\pi\pi}^2k_\rho^3}{6\pi m_\rho^2},
	\label{eq:gGamma}
\end{equation}
with $k_\rho^3$ accounting for the broadening of the resonance as
phase space opens up.
Thus, one could determine the physical $\rho$ meson width by 
extrapolating $g_{\rho\pi\pi}$ to physical quark mass, and then using
Eq.~(\ref{eq:gGamma}).

The relationship between the energy levels of scattering states and
phase shifts extends to weak non-leptonic decays, again in the
elastic region.\cite{Lellouch:2000pv}
Thus, it is not a conceptual problem to calculate the phase shifts in 
$K$ decays.
Unfortunately, for for inelastic decays, such as $B$ meson decays, 
it remains an unsolved problem.

Often an even bleaker picture is painted, based on a superficial
understanding a theorem of Maiani and Testa.\cite{Maiani:1990ca}
The theorem assumes an infinite volume and is, thus, relevant only to 
extremely large volumes, where there the level spacing is so small 
that it is not possible, because of statistical errors, to resolve 
them.
Then one simply observes that Euclidean correlation functions are 
real, so phases cannot appear without analytic continuation to 
Minkowski space.
Although possible in principle,\cite{GlimmJaffe} this is again not 
practical when the correlation functions are computed only at 
discrete values of time separation.
As with many theorems, the way out is to relax one of the hypotheses:
in this case, the assumption of infinite volume.

Although the study of finite volume effects, in the elastic region, is 
on a sound conceptual footing, there are several technical challenges.
First, as mentioned above, the quark masses must be somewhat 
smaller than usually used.
Second, many independent Monte Carlo runs, with varying $L$ but other 
parameters fixed, are needed.
Third, the quenched approximation is not suited to these calculations 
at all, because the would-be $\eta'$ produces unphysical terms in 
the~$1/L$ expansion.\cite{Bernard:1996ps}
Finally, small statistical errors are needed, to trace out the 
level-crossing regions, especially when the resonance is narrow.
Further practical aspects have been considered by
Lin et al.\cite{Lin:2001ek}

\section{Top Ten Trends}
\label{sec:ten}

When the Editor invited me to contribute a chapter for the Handbook of
QCD, one possibility that was raised was a list of the top ten results
from lattice QCD.
One version of this idea, which was quickly discarded, would have been 
a list of specific numerical results.
Such lists are quickly out of date, and are better obtained from
recent from lattice conferences.

Nevertheless, it is tempting to list some of the main trends.
In contrast to the rest of the chapter, which focused on practical
aspects, the following topics should enjoy broad interest and influence
one's thinking about QCD.
\begin{enumerate}

\item {\em Hadron spectrum:}
{\em Ab initio} calculations of the hadron spectrum were one of the
early motivations for numerical lattice QCD.
The results of a recent calculation\cite{AliKhan:2001tx} are shown
in Fig.~\ref{fig:hadrons}, comparing quenched and $n_f=2$ calculations.
\begin{figure}
	\centering
	\includegraphics[width=0.8\textwidth]{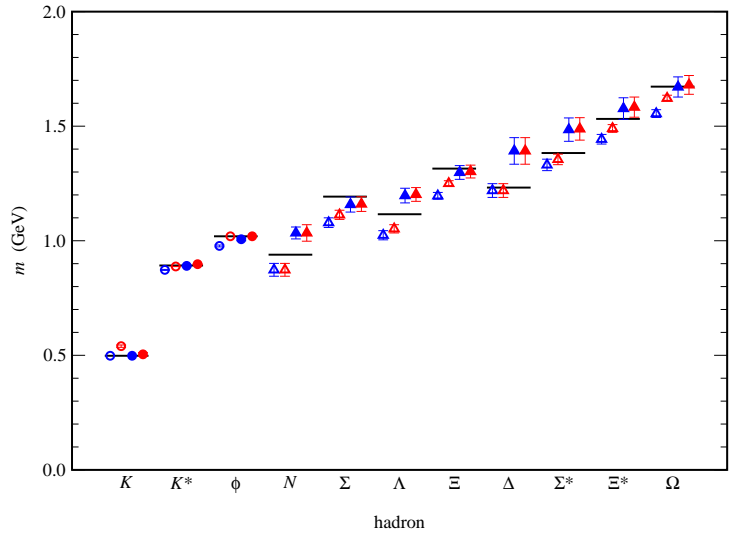}
	\caption[fig:hadrons]{The hadron spectrum.\cite{AliKhan:2001tx}
	Solid (open) point show the results of unquenched (quenched) lattice
	QCD.
	Blue (red) points tune the strange quark's mass to the $K$ ($\phi$).
	}
	\label{fig:hadrons}
\end{figure}
The quenched approximation does surprisingly well, agreeing with
experiment within several percent.
For hadrons with strange quark, particularly the $\phi$ and the
$\Omega^-$, the unquenched calculations agree even better.
For hadrons with only up and down quarks, for example the nucleon,
it remains difficult to control finite-size effects.

\item {\em Glueballs must exist:}
When QCD was first put forward, it seemed plausible, but not obvious,
that gluons would bind together to form hadrons that lie beyond the
classification scheme of the quark model.
Such states are called glueballs and hybrids.
Although there is still controversy over whether such glueballs have
been observed in the laboratory, lattice calculations give strong
evidence that their existence is a prediction of~QCD.
There is even a well-argued case that the $f_0(1710)$ is the
lowest-lying scalar glueball, based on calculations of the masses and 
coupling to pseudoscalar meson pairs.\cite{Sexton:1995kd}

\item {\em Confinement explained:}
An early result of the strong coupling expansion of lattice gauge
theory was the so-called ``area law'' for Wilson loops.
A Wilson loop is the line integral of the gauge potential around a
closed curve ${\cal C}$, or
$W({\cal C})={\rm P}\exp\left(\int_{\cal C}A\cdot dx\right)$, 
where ${\rm P}$ denotes path ordering.
For a $R\times T$ rectangle, a short calculation in the strong coupling
limit shows\cite{Wilson:1974sk}
\begin{equation}
	W(R,T)\sim \left(\frac{1}{g_0^2}\right)^{RT},
	\label{eq:RT}
\end{equation}
namely, that $\ln W({\cal C})$ is proportional to the area enclosed
by~${\cal C}$.
For large $T$, $\ln W(R,T)\sim -V(R)T$, where $V(R)$ is the potential
energy of the gluon field between two static sources of color.
Thus, the area law shows a linearly rising potential, which confines
quarks.
It is amazing that confinement emerges so simply.

Numerical calculations confirm that the linear rise persists for
non-Abelian gauge theories into the weak (bare) coupling regime,
{\em i.e.}, towards the continuum limit.
(It goes away for the Abelian U(1) lattice gauge theory.)
A~recent survey is in the review of Bali.\cite{Bali:2000zv}
The lattice calculations trace out unmistakably a potential
that is Coulombic at short distances (as expected from asymptotic
freedom) and linear at large distances (as needed for confinement).

Insight into confinement is not limited to the heavy-quark potential.
From a chromoelectric perspective, studies of the gluonic flux between
static sources are starting to paint an even more detailed
picture.\cite{Morningstar:1998da}
From a complementary perspective, several ideas for the condensation
of chromomagnetic monopoles have been studied.\cite{Frohlich:2001eq}

\item {\em NRQCD and HQET:}
It is worth recalling that both non-relativistic QCD (NRQCD) and
heavy-quark effective theory (HQET) were invented, in part, to
facilitate lattice calculations of heavy-quark
systems.\cite{Lepage:1987gg,Eichten:1987xu}
With continuum ultraviolet regulators, instead of a lattice, these
effective field theories have become powerful tools for studying,
for example, quarkonium production in high-energy collisions, or
semi-leptonic decays of $B$ mesons.
They are important for understanding the CKM matrix,
and QCD itself.

\item {\em The light quarks are very light:}
There is now considerable evidence from lattice QCD that the
strange quark has a mass lower than previously thought.
The up and down quarks are correspondingly lighter, 
as expected from chiral symmetry:
$(m_u+m_d)/2m_s\approx m_\pi^2/2m_K^2\approx1/25$.
A recent calculation with dynamical quark loops
finds the $\overline{\rm MS}$ mass\cite{AliKhan:2001tx}
\begin{equation}
	\bar{m}_s(2~{\rm GeV}) = 88^{+4}_{-6}~{\rm MeV}.
\end{equation}
This was not an easy conclusion to reach, because lattice spacing
effects and quenching effects are both large.
If either is not taken into account, a larger estimate of $\bar{m}_s$ is
obtained.
One implication of this result is that very little of the mass of
everyday objects comes from the Higgs mechanism of the electroweak
theory.
Almost all of it comes from gluons.

\item {\em QCD has a phase transition:}
There are theoretical grounds to expect that pure gauge theory
has a deconfinement transition, and that in QCD the high-temperature
phase restores chiral symmetry.
(See Part~8 of Vol.~3 of the Handbook.)
But without numerical lattice QCD it is impossible to compute the
critical temperature of the phase transition(s).
The calculation of the pure gauge transition temperature was one of the
first large-scale numerical simulations to yield, persuasively, a
physically interesting result that was not known from experiment.

\item {\em $\alpha_s$ from hadrons agrees with $\alpha_s$ from
high-energy scattering:}
One of the entries for $\alpha_s$ in the Review of Particle
Properties\cite{Groom:2000in} comes from lattice calculations of
charmonium and bottomonium spectra.
It lies within the range of the other determinations, which come from
high-energy scattering.
The agreement shows that the QCD of confinement is also the QCD of
partons: QCD works at long and short distances.

\item {\em Determination of the CKM matrix:}
Lattice calculations of matrix elements in $B$, $D$, and $K$ decays
are now seen as essential to testing the standard CKM picture of flavor
and $CP$ violation.\cite{Beneke:2002ks}
The most appealing aspect of the CKM mechanism is that it is very
predictive, relating many $CP$ conserving and $CP$ violating processes
back to four parameters, only one of which violates~$CP$.
With measurements of $CP$ conserving processes, such as semi-leptonic 
decays and neutral meson mixing, and calculations of the corresponding 
form factors and mixing amplitudes, one can predict the strength of 
$CP$ violation.
Once unquenched QCD calculations of these quantities mature, this area
could well become the most important contribution of lattice~QCD.

\item {\em Lattice chiral symmetry:}
The last few years have witnessed remarkable progress of understanding
chiral symmetry in lattice gauge
theory.\cite{Niedermayer:1998bi,Neuberger:1999ry,Luscher:1999mt}
In the context of QCD, the contribution is significant:
a tool for more rigorous derivations of well-known soft-pion
theorems,\cite{Chandrasekharan:1998wg}
or an ingredient in calculations of kaon matrix elements relevant to
CKM phenomenology.\cite{Blum:1996jf}
The contribution of this development of lattice gauge theory to
electroweak physics and to extensions of the Standard Model promises to
be much greater.

\item {\em Perturbation theory is universal:}
A somewhat underappreciated, but nevertheless important, achievement
is the proof that perturbation theory is universal.\cite{Reisz:1988kk}
After renormalization, the $a\to0$ limit of perturbative QCD is the
same for all lattice actions.
Since lattice field theory maintains rigorous control over the
ultraviolet at all steps of the analysis, it puts more familiar (i.e.,
continuum) ultraviolet regulators on a sounder footing, because they
are all connected to lattice gauge theory through BPHZ renormalization.
\end{enumerate}

\section{Summary: Computational vs.\ Theoretical Physics}

Lattice gauge theory is a broad subject, with many applications beyond
QCD, especially now that chiral fermions are better understood.
The lattice provides a well-defined mathematical foundation for quantum
field theory, which remains one of its strongest attractions.
In particular, it permits, in least in principle, real non-perturbative
calculations.

Especially in the realm of QCD, one technique for calculation is most
prominent, namely computing the functional integral numerically, via
Monte Carlo methods.
Because we live in 3+1 dimensional spacetime, and because real hadrons
have several scales ($m_q$, $\Lambda$, $m_Q$), the problem is far from
easy.
When discussing numerical lattice QCD, usually only the
computational aspects that are emphasized, perhaps overemphasized.

Computing is necessary, but not sufficient.
In addition to the physical scales of hadrons, the numerical technique
has an ultraviolet cutoff (from the lattice) and an infrared cutoff
(from a finite spacetime volume).
A~whole branch of theoretical physics has been developed to understand
the cutoff effects, and how to get the most out of scarce computer
resources.
Similarly, for practical reasons one must cope with light quark masses
that are not so light, and heavy quark masses near the ultraviolet
cutoff.
Here tools are available from continuum QCD, but they require some
redesign to be usefully deployed.

Even if the numerical solution of QCD were accessible solely by brute
force, it would be wasteful to rely on brute force alone.
An example, discussed at the end of Sec.~\ref{sec:sym}, is the control
of cutoff effects.
It is much more efficient to run at several lattice spacings than to put
all computer resources onto the finest conceivable lattice.
A~similar comment applies to the light quark mass, but here, at least,
it is completely routine to do~so.

In the past decade computing has been essential to test several new
ideas, such as faster algorithms, methods for heavy quarks, and
non-perturbative renormalization.
Most of the progress in lattice QCD has come, however, from the ideas
and not from Moore's Law, which observes that computer power per dollar
doubles every 18 months.
Because we live in four dimensions, and because the algorithms slow
down, one can double the lattice spacing (at fixed computer cost) only
every 6--8 years.
Increases in computing are essential, but not sufficient.

It is likely that this trend will continue.
For example, non-perturbative renormalization is helpful for light
quarks, but remains immature for heavy quarks.
For kaon physics relatively new, chirally symmetric, methods for the
quarks are promising.
But they are even more computationally intensive.
One of the methods, domain wall fermions, adds a fifth dimension to the
lattice and to the scaling laws in Sec.~\ref{sec:why}.
The other methods are similarly difficult.
Thus, especially here, the need for ideas to reduce the computational
burden (for fixed error bar) is great.

Even if no new ideas emerge plenty of traditional theoretical physics is
needed to understand numerical lattice gauge theory and, therefore, QCD.
Because effective field theories provide tools to reduce the
uncertainties, there is always more work to do:
calculating the next order in chiral perturbation theory,
or programming the next set of operators to reduce discretization
effects.
This work is necessary, especially when the results of lattice QCD are
needed to understand flavor or collider physics.
One should assume, however, the there will be new ideas, and that the
theoretical and computational sides will continue to inform each other.


\section*{Acknowledgments}
I would like to thank Misha Shifman for the invitation to write this
chapter, and for his patience as it was being written.
Over the years, many people have influenced my views on lattice gauge
theory, particularly Peter Lepage, Martin L\"uscher, and Kenneth Wilson.
My collaborators in developing a theory of heavy-quark discretization
effects, Aida El-Khadra, Shoji Hashimoto, Paul Mackenzie, and Tetsuya
Onogi, have had a significant influence on my way of thinking.
While writing this chapter, I benefitted from a conversation on chiral
perturbation theory with Maarten Golterman and Stephen Sharpe.
Finally, I would like to thank the Aspen Center for Physics and
Zoltan Ligeti for providing an opportunity to rehearse the material
before a live (and lively!) audience.
Fermilab is operated by Universities Research Association Inc.,
under contract with the U.S.\ Department of Energy.

\end{document}